\documentclass[aps,prd,twocolumn,superscriptaddress]{revtex4-1} 

\RequirePackage{graphicx}
\RequirePackage[colorlinks,citecolor=blue,urlcolor=blue,linkcolor=blue]{hyperref}

\usepackage{enumerate}
\usepackage{amsmath}
\usepackage{amsfonts}
\usepackage{amssymb}
\usepackage{color}

\RequirePackage[capitalize]{cleveref}
\crefname{figure}{fig.}{figs.}
\crefname{table}{table}{tables.}
\crefname{equation}{eqn.}{eqns.}
\crefname{section}{sec.}{secs.}

\graphicspath{{figs/}{./}} 

\begin{document}

\title{MeV-scale performance of water-based and pure liquid scintillator detectors}

\author{B. J. ~Land}
\affiliation{University of Pennsylvania, Philadelphia, PA, USA}
\affiliation{University of California, Berkeley, CA 94720-7300, USA}
\affiliation{Lawrence Berkeley National Laboratory, CA 94720-8153, USA}
\author{Z. ~Bagdasarian}
\affiliation{University of California, Berkeley, CA 94720-7300, USA}
\affiliation{Lawrence Berkeley National Laboratory, CA 94720-8153, USA}
\author{J. ~Caravaca}
\affiliation{University of California, Berkeley, CA 94720-7300, USA}
\affiliation{Lawrence Berkeley National Laboratory, CA 94720-8153, USA}
\author{M. Smiley}
\affiliation{University of California, Berkeley, CA 94720-7300, USA}
\affiliation{Lawrence Berkeley National Laboratory, CA 94720-8153, USA}
\author{M.~Yeh}
\affiliation{Brookhaven National Laboratory, Upton, NY 11973-500, USA}
\author{G. D. ~Orebi Gann}
\affiliation{University of California, Berkeley, CA 94720-7300, USA}
\affiliation{Lawrence Berkeley National Laboratory, CA 94720-8153, USA}

\begin{abstract}
This paper presents studies of the performance of water-based liquid scintillator in both 1-kt and 50-kt detectors.
Performance is evaluated in comparison to both pure water Cherenkov detectors and a nominal model for pure scintillator detectors.  
Performance metrics include energy, vertex, and angular resolution, along with a metric for ability to separate the Cherenkov from the scintillation signal, as being representative of various particle identification capabilities that depend on the Cherenkov / scintillation ratio.
We also modify the time profile of scintillation light to study the same performance metrics as a function of rise and decay time.  
We go on to interpret these results in terms of their impact on certain physics goals, such as solar neutrinos and the search for Majorana neutrinos. This work supports and validates previous results, and the assumptions made therein, by using a more complete detector model and full reconstruction. We confirm that a high-coverage, 50-kt detector would be capable of better than 10 (1)\% precision on the CNO neutrino flux with a WbLS (pure LS) target in 5 years of data taking. A 1-kt LS detector, with a conservative 50\% fiducial volume of 500~t, can achieve a better than 5\% detection.  Using the liquid scintillator model, we find a sensitivity into the normal hierarchy region for Majorana neutrinos, with half life sensitivity of $T^{0\nu\beta\beta}_{1/2} > 1.4 \times 10^{28}$ years at 90\% CL for 10 years of data taking with a Te-loaded target.  
\end{abstract}

\maketitle

\section{\label{sec:intro} Introduction}

These are exciting times for neutrino physics, with a number of open questions that can be addressed by next-generation detectors.
Advances in technology and innovative approaches to detector design can drive the scientific reach of these experiments.
A hybrid optical neutrino detector, capable of leveraging both Cherenkov and scintillation signals, offers many potential benefits.  The high photon yield of scintillators offers good resolution and low thresholds, while a clean Cherenkov signal offers ring imaging at high energy, and direction resolution at low energy.  The ratio of the two components provides an additional handle for particle identification that can be used to discriminate background events.  

There is significant effort in the community to develop this technology, including 
target material development~\cite{wbls,Bignell_2015,Buck_2016,bnl_slow_scintillator,steve_slow_scintillator,Graham_2019,Aberle_2013,doi:10.1063/1.3112609,Ashenfelter_2019,CUMMING20191,bay2020optimization,doi:10.1063/1.4927458}, 
demonstrations of Cherenkov light detection from scintillating media~\cite{chess_epjc,chess_wbls,Gruszko_2019,Li:2015phc}, 
demonstrations of spectral sorting~\cite{dichroicon,dichroicon2}, 
fast and high precision photon detector development~\cite{Lyashenko2020,Adams:2015kkx,Ramberg:2010zza,Oberla:2015oha,Siegmund:2011wzq,Barros:2015pjt,PMT2012,PhysRevD.97.052006},
complementary development of reconstruction methods and particle identification techniques~\cite{Wonsak:2018uby,Wonsak_Dresden,Dunger:2019dfo,Aberle:2014,elagin,Jiang:2019cnb,Li:2018}, and development of a practical purification system at UC Davis.

One approach to achieving a hybrid detector is to deploy water-based liquid scintillator (WbLS)~\cite{wbls}, a novel target medium that combines water with pure organic scintillator, thus leveraging the benefits of both scintillation and Cherenkov signals in a single detection medium, with the advantage of high optical transparency and, thus, good light collection.
Many experiments are pursuing this technology for a range of applications, including a potential ton-scale deployment at ANNIE at FNAL~\cite{annie,annie_eof,anghel2015letter}, possible kt-scale deployments at the Advanced Instrumentation Testbed (AIT) facility in the UK~\cite{AIT,watchman,danielson2019directionally} and in Korea~\cite{KoreaWbLS}, and, ultimately, a large (25--100 kt) detector at the Long Baseline Neutrino Facility, called \textsc{Theia}.  The \textsc{Theia} program builds heavily on early developments by the LENA collaboration~\cite{WURM2012685}.  Such a detector could achieve an incredibly broad program of neutrino and rare event physics, including highly competitive sensitivity to long-baseline neutrino studies, astrophysical searches, and even scope to reach into the normal hierarchy regime for neutrinoless double beta decay~\cite{theia_wp,asdc,theia_richie,steve}.  

In this paper, we study the low-energy performance of such a detector for a range of different target materials, and compare the results to that for a pure water Cherenkov detector, and a pure liquid scintillator (LS) detector, using linear alkyl benzene (LAB) with 2 g/L of the fluor 2,5-Diphenyloxazole (PPO) as the baseline for comparison.  The goal of this paper is to contrast WbLS performance to LS under similar assumptions, and to validate the simple model used in~\cite{theia_wp} with a more complete optical model, more detailed detector simulation, and full event reconstruction.  

Properties for the pure LS detector are taken from measurements by the SNO+ collaboration~\cite{snoplus,snoplus_private}.  We start by considering three WbLS target materials.  Each cocktail is a combination of water with liquid scintillator, with differing fractions of the organic component: 1, 5 and 10\% concentration by mass.   
WbLS properties are based on bench-top measurements~\cite{drew_wbls,chess_wbls} or evaluated based on constituent components, as described in \Cref{sec:model}.
Measurements of these WbLS materials demonstrated a very fast timing response: with a rise time consistent with 0.1~ns, and a prompt decay time on the order of 2.5~ns.  Since this fast time profile increases the overlap between the prompt Cherenkov and delayed scintillation signals, we also consider materials in which we delay the scintillation time profile by some defined amount, to study the impact of a ``slow scintillator'', for both  pure LS and  WbLS.  Such materials are under active development~\cite{bnl_slow_scintillator,steve_slow_scintillator}.

It should be noted that, throughout this article, the pure LS in question is LAB + 2g/L PPO, and the LS component of the WbLS under consideration is formulated from these constituent materials, with additional surfactants and other components to achieve stability, good light yield, and good attenuation properties.  Any comparisons made are specific to these materials.  Further optimization is likely possible, resulting in further improvements to performance, such as use of a secondary fluor to shift the emitted spectrum.  While we consider materials with a delayed time profile, in order to understand the impact of improved separation of the Cherenkov component, these models are hypothetical, and intended to motivate further material development.

Metrics used for these performance studies include the energy resolution (dominated by photon counting and quenching effects), vertex resolution, direction resolution, and a statistic chosen to represent the separability of the Cherenkov and scintillation signals.  This is representative of low-energy performance capabilities such as particle identification, which may rely on separating the two populations.  The final choice of a detector material for any particular detector would depend on the physics goals, which will place different requirements on each aspect of detector performance.  In all cases, we focus on the low-energy regime.  Performance studies at the high energies relevant for neutrino beam physics are underway, and will depend on a different combination of factors, so may yield different optimizations.

We consider both a 1-kt and a 50-kt total mass detector, as being representative of experiments currently under consideration.  It should be noted that the 1-kt detector results in a small, 500-ton fiducial volume for the physics cases under study.
The metrics presented in this paper are highly dependent on the transit time spread (TTS) of the photodetectors, and we present results for four hypothetical photodetectors models in this study.
To span the range of available options, our four models have TTS of 1.6~ns, 1.0~ns, 500~ps, and 70~ps (sigma).
In each case we assume 90\% coverage, with a constant representative quantum efficiency (QE) used for all four models.

To understand the reach of the detector capabilities studied here, we discuss the impact for several low-energy physics goals, in particular considering scope for a precision measurement of CNO solar neutrinos, and normal hierarchy sensitivity for neutrinoless double beta decay (NLDBD)~\cite{steve,theia_wp}. 
Large-scale scintillator detectors such as Borexino~\cite{Agostini} and KamLAND-Zen~\cite{KamLAND-Zen} are leaders in the fields of solar neutrinos and searches for NLDBD, respectively, and new scintillator detectors such as SNO+~\cite{snoplus} and JUNO~\cite{Ge:2015bfa,juno} are taking data or under construction.  There is much interest in the community in using new solar neutrino data for precision understanding of neutrino properties and behavior, as well as for solar physics~\cite{PhysRevD.101.123031}.
The proposed \textsc{Theia} experiment has discussed and evaluated the potential of a multi-kiloton, high-coverage WbLS detector for the purposes of solar neutrino detection and NLDBD~\cite{theia_richie,theia_wp}, where the latter would deploy inner containment for an isotope-loaded pure LS target, adapting techniques from SNO+ and KamLAND-Zen.  Studies such as those presented here can help to inform future detector design.

\Cref{sec:model} presents details of the scintillator model used.
\Cref{sec:methods} describes the simulation and analysis methods, including the reconstruction algorithms applied. 
\Cref{sec:impact} presents results for performance of the measured WbLS cocktails, including photon counting and reconstruction capabilities.
\Cref{sec:timing-impact} presents the results as a function of rise and decay time, considering both the pure LS and a 10\% WbLS.
\Cref{sec:physics} discusses these performance results in light of their impact on certain selected physics goals, and \Cref{sec:conclusions} concludes.

\section{\label{sec:model} Water-based Liquid Scintillator model}

For Monte Carlo simulation of photon creation and propagation in WbLS, we use the  Geant4-based~\cite{geant4} RAT-PAC framework~\cite{ratpac}. 
Cherenkov photon production is handled by the default Geant4 model, G4Cerenkov. Rayleigh scattering process is implemented by the module developed by the SNO+ collaboration~\cite{snoplus_private}.
The GLG4Scint model 
handles the generation of scintillation light, as well as photon absorption and reemission. We utilize the light yield as measured in WbLS (1\%,5\%, and 10\% solutions) in Ref.~\cite{chess_wbls}, and scintillation emission spectrum and time profile as taken from Ref.~\cite{drew_wbls}.  These time profile measurements were confirmed with both x-ray excitation~\cite{drew_wbls} and direct measurements with $\beta$ and $\gamma$ sources~\cite{chess_wbls}.  
Other inputs to the water-based liquid scintillator optical model have either not yet been measured directly, or measurements (such as those in~\cite{wbls}) are of early prototypes, and do not represent recent developments of these materials.
To create an optical model of WbLS materials, those inputs are estimated as a combination of those from water and pure LS, as described below.

\subsection{Refractive index estimation}

In order to estimate the refractive index for WbLS, $n$, we use Newton's formula for the refractive index of liquid mixtures \cite{newton_rindex}:
\begin{equation}
n = \sqrt{\phi^{}_{labppo} n^2_{labppo} + \phi^{}_{water} n^2_{water}},
\end{equation}
where $\phi$ denotes the volume fraction of a corresponding component, while $n_{labppo}$ and $n_{water}$ correspond to the measured refractive indexes for pure LS \cite{snoplus_private} and water \cite{rindex_water} as a function of wavelength. 
At 400 nm, the refractive index of water is 1.344, and 1.505 for the pure LS. 
The estimates for 1\%, 5\%, and 10\% WbLS at 400 nm are 1.347, 1.359, and 1.372, respectively. The full wavelength dependence is included in the simulation.
Due to the dominant fraction of water, the WbLS refractive index is very similar to that of pure water.

\subsection{Absorption and scintillation reemission}

The absorption coefficient, $\alpha$, of WbLS depends on the molar concentration, $c$, of each of the components as:
\begin{equation}
\alpha (\omega)= c_{lab}\epsilon_{lab}(\omega) + c_{ppo} \epsilon_{ppo}(\omega) +  c_{water} \epsilon_{water}(\omega),
\end{equation}
where $\epsilon_{lab}$, $\epsilon_{ppo}$ and $\epsilon_{water}$ are the molar absorption coefficients of LAB, PPO~\cite{snoplus_private}, and water (taken from Ref.~\cite{alength_water1} for wavelengths over 380~nm and from Ref.~\cite{alength_water2} for wavelengths below 380~nm).

A photon absorbed by the scintillator volume has a non-zero probability of being reemitted. 
This reemission process becomes important at low wavelengths where the absorption by scintillator is dominant. 
As a result, photons are shifted to longer wavelengths where the detection probability is higher due to a smaller photon absorption and a greater PMT quantum efficiency. 
The probability $p^{reem}_{i}$ of a component $i$ absorbing a photon of frequency $\omega$ is determined as the contribution of the given component to the total WbLS absorption coefficient:
\begin{equation}
p^{reem}_{i}(\omega)	=  \phi_i\alpha_i(\omega) / \alpha(\omega),
\end{equation}
where $\phi_{i}$ is the volume fraction of component $i$ in WbLS. After a photon is absorbed, it can be reemitted with a $59\%$ probability for LAB and an $80\%$ probability for PPO \cite{snoplus_private}, following the primary emission spectrum.

\subsection{Scattering length}

The Rayleigh scattering length, $\lambda^{s}$, is estimated for WbLS as:
\begin{equation}
\lambda^{s}(\omega) =\left(\phi^{}_{lab}\lambda_{lab}^{-1}(\omega) +\phi^{}_{water}\lambda_{water}^{-1}(\omega)\right)^{-1},
\label{eq:scat}
\end{equation}
where $\lambda_{lab}$ and $\lambda_{water}$ are the scattering lengths for LAB and water, respectively, both taken from \cite{snoplus_private}. It was noted that the addition of PPO does not change $\lambda^s$ and thus it is omitted in  Eq.~\ref{eq:scat}.

The resulting values of both absorption and scattering lengths for WbLS are close to those of pure water. It is known that this method overestimates the attenuation lengths, in particular, the scattering, given the complex chemical structure and composition of WbLS. A long-arm measurement of WbLS absorption and scattering lengths is planned in the near future.  However, recent (unpublished) data from BNL demonstrate scattering lengths on the scale of the largest size of detector being considered here.  Thus the known simplification is considered an acceptable approximation until further data becomes available.


\section{\label{sec:methods} Simulation and analysis methods}

The WbLS models developed in~\cite{chess_prc}, and described above, can be used to evaluate the performance of these materials in various simulated configurations.
Of interest are large, next-generation detectors such as \textsc{Theia}~\cite{theia_wp}, which could contain tens of kilotons of target material instrumented with high quantum efficiency photodetectors at high coverage, and proposed detectors in the range of one to a few kt, such as AIT~\cite{AIT}.
To evaluate these materials, two detector configurations are simulated: a 1-kt detector and a 50-kt detector, both with 90\% coverage of photon detectors as a baseline.
The different concentration WbLS materials studied in~\cite{chess_wbls},  1\%,  5\% and 10\% WbLS, are simulated and compared to both water and pure (100\%) scintillator material~\cite{snoplus_private}.

\subsection{Monte Carlo simulation}

Fully simulating next-generation detector sizes instrumented with 3D models of photon detectors at the desired coverage of 90\% requires significant computational resources.
This is especially true when studying multiple geometries, as the simulation typically must be rerun for each geometry.
To avoid this redundancy, RAT-PAC~\cite{ratpac} can easily simulate a sufficiently large volume of material and export the photon tracks to an offline geometry and photon detection simulation.
Using this method, 2.6-MeV electrons are simulated at the center of a large volume of target material, isotropic in direction, and the resulting tracks are stored for later processing by a detector geometry model and a photon detector model.  This energy is chosen as being representative of a number of low-energy events of interest, including reactor antineutrinos, low-energy solar neutrinos, and the end-point of double beta decay for both $^{136}$Xe and $^{130}$Te.

\subsubsection{Detector geometry}
Each detector configuration is modeled as a right cylinder with diameter and height of 10.4~m and 38~m for the 1-kt and 50-kt sizes, respectively.  Specifically, this calculation achieves a 1-kt and 50-kt total mass for the pure LS detector, with slightly modified target masses for the other target materials, based on different densities (the LS under consideration has a density of 0.867 g/cm$^3$, while WbLS is within a few percent of 1.0 g/cm$^3$).
The photon tracks from stored events that are found to intersect with the cylinder representing the detector boundary are stored as potential detected photons (``hits'') for each event.
In this way, the boundary of each active volume acts as a photon-detecting surface that provides all information about each photon to a photon detector model. 

This simulation approach ignores several effects present in real detectors, including reflections off of the photodetectors, position uncertainty due to photodetectors size, and false-positive photon detection (noise) from real photodetectors.
Typically, reflected photons will have a much longer path length than non-reflected photons, arrive much later, and add little information to event reconstruction, so a lack of photodetector reflections will have minimal impact on the metrics presented.
Particularly, for angular reconstruction, we exclude all but the most-prompt photons, further reducing potential impact of reflections.
The impact of position resolution was explored here by randomly shifting the position of detected photons by up to 100~cm, and studying the impact on the reconstruction metrics shown later in the paper.
Ultimately, no statistically significant change was observed after smearing the photon detection positions, which can be understood by noting that the photon detection positions are far from the center-generated events studied here.
In the 1-kt (50-kt) detector, this smearing results in (at most) an 11~deg (3~deg) shift in the photon position, which is well below the best angular resolution achieved in this study.
This indicates that position uncertainty of real photodetectors will have minimal impact on results provided.
As a consequence of this, no reliance is made on the purported position resolution of LAPPDs, and we assume they could be deployed as devices with single-anode readout, similar to a PMT, which report only the time of the photon arrival.
Finally, noise in the detector is expected to be sub-dominant to actual scintillation light, however it may be significant compared to Cherenkov light, depending on the size of the time window used to select events. 
As will be shown, Cherenkov photons are selected from tight time windows on the order of 1~ns, meaning a total noise rate of order 1~GHz would be necessary to expect one noise photon within the Cherenkov window.
In the 50-kt detector, between 10,000 and 100,000 photodetectors (depending on the exact form factor used) would be necessary to achieve the desired coverage, which places an approximate upper limit on the per-photodetector noise rate of 10~kHz.
This is an acceptable upper bound compared to modern PMTs~\cite{hamamatsu_catalog}, so ignoring noise is considered to be a reasonable approximation and should have little impact on the results presented.

\subsubsection{Photon detection}
Photon detectors vary in their probability of detecting a photon as a function of wavelength (the QE) and their time resolution (TTS).
Recently developed prototype photomultiplier tubes (PMTs) like the R5912-MOD~\cite{r5912mod} can achieve a TTS of 640~ps (sigma), while commercially available large-area PMTs like the R7081-100 or R5912-100~\cite{hamamatsu_catalog} are quoted at a TTS of 1.5~ns or 1.0~ns (sigma), which may be better (worse) at higher (lower) bias voltage.
Next generation photodetectors such as large-area picosecond photon detectors (LAPPDs)~\cite{lappd} achieving a TTS of 70~ps (sigma).
Four hypothetical photon detector models are considered for each material and geometry, to span this range: 
\begin{enumerate}
\item \textit{``PMT''} a generic commercially available large-area high-QE PMT, similar to an R5912-100 or R7081-100~\cite{hamamatsu_catalog}, with 34\% peak QE and 1.6-ns TTS (sigma). 
\item \textit{``FastPMT''} a hypothetical PMT with a similar QE but smaller TTS of 1.0~ns (sigma).
\item \textit{``FasterPMT''} a hypothetical PMT again with a similar QE but even smaller TTS of 500~ps (sigma).
\item \textit{``LAPPD''} a next-generation device such as a large-area picosecond photodetector (LAPPD)~\cite{lappd} with similar QE but a 70-ps TTS (sigma).
\end{enumerate}
The same QE is used for all four models, assuming that future LAPPDs can reach comparable QE to existing Hamamatsu large-area PMTs.

\begin{figure}
\centering
\includegraphics[width=\columnwidth]{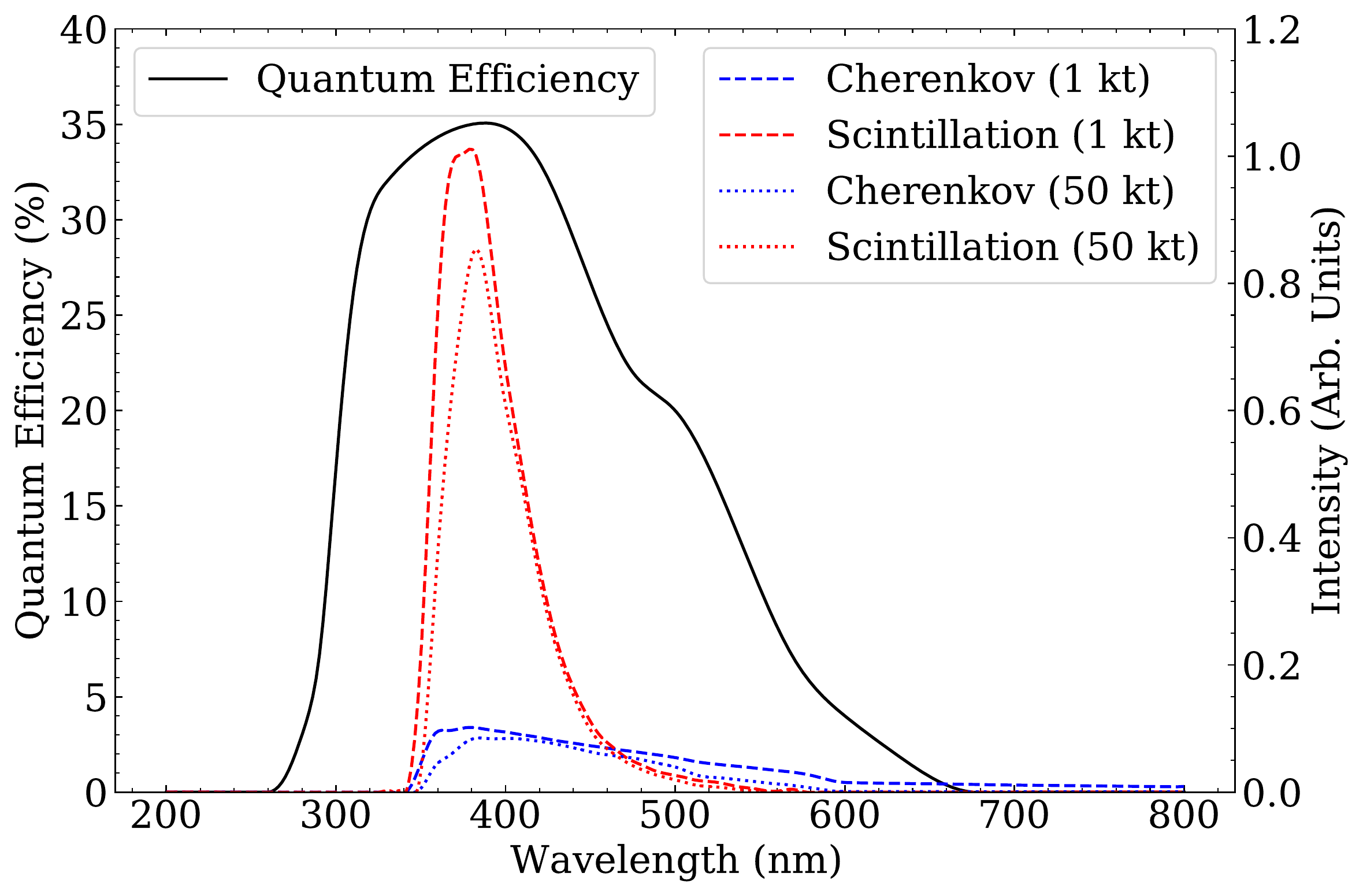}
\caption{\label{fig:qe} The quantum efficiency (QE) used for photon detector models considered here (digitized from~\cite{hamamatsu_catalog}). Also shown are Cherenkov and scintillation photon spectra for center-generated events in the  1\% WbLS material in the two detector sizes prior to application of QE, i.e. including all optical effects, but no photon detection effects. The relative normalization of the spectra have been preserved, with the maximum value normalized to 1.0.}
\end{figure}

A coverage of 90\% using these devices is simulated by accepting only 90\% of potential hits for the event.
This high coverage is chosen as being slightly less than the maximum packing of identical circles on a plane: 90.7\%.
For a square device like an LAPPD, or a mixture of dissimilar sized devices, higher coverage may be achievable.
The QE is accounted for by randomly accepting hits according to the value of the QE curve (shown in \Cref{fig:qe} with typical wavelength spectra) at the wavelength of the hit.
For the selected hits, the intersection position with the geometry model is taken as the detected position.
Finally, a normally distributed random number with a width corresponding to the TTS of the photon detector model is added to the truth time of the hit to get the detected time.
These detected hit position and times can then be passed to reconstruction algorithms for further analysis.

\subsection{Event reconstruction}

To evaluate the performance of the different materials under different detector configurations, a fitter was developed to reconstruct the initial vertex parameters based on detected hit information.
Position and time reconstruction are both aided by the large number of isotropic scintillation photons, while direction reconstruction relies on identification of non-isotropic Cherenkov photons.
As Cherenkov photons are prompt with respect to scintillation photons, the reconstruction will first identify prompt photons, and then use them to reconstruct direction in a staged approach.
Promptness is defined in terms of the hit time residual $t_{resid}$ distribution.

\begin{figure*}[]
\centering
\includegraphics[width=\columnwidth]{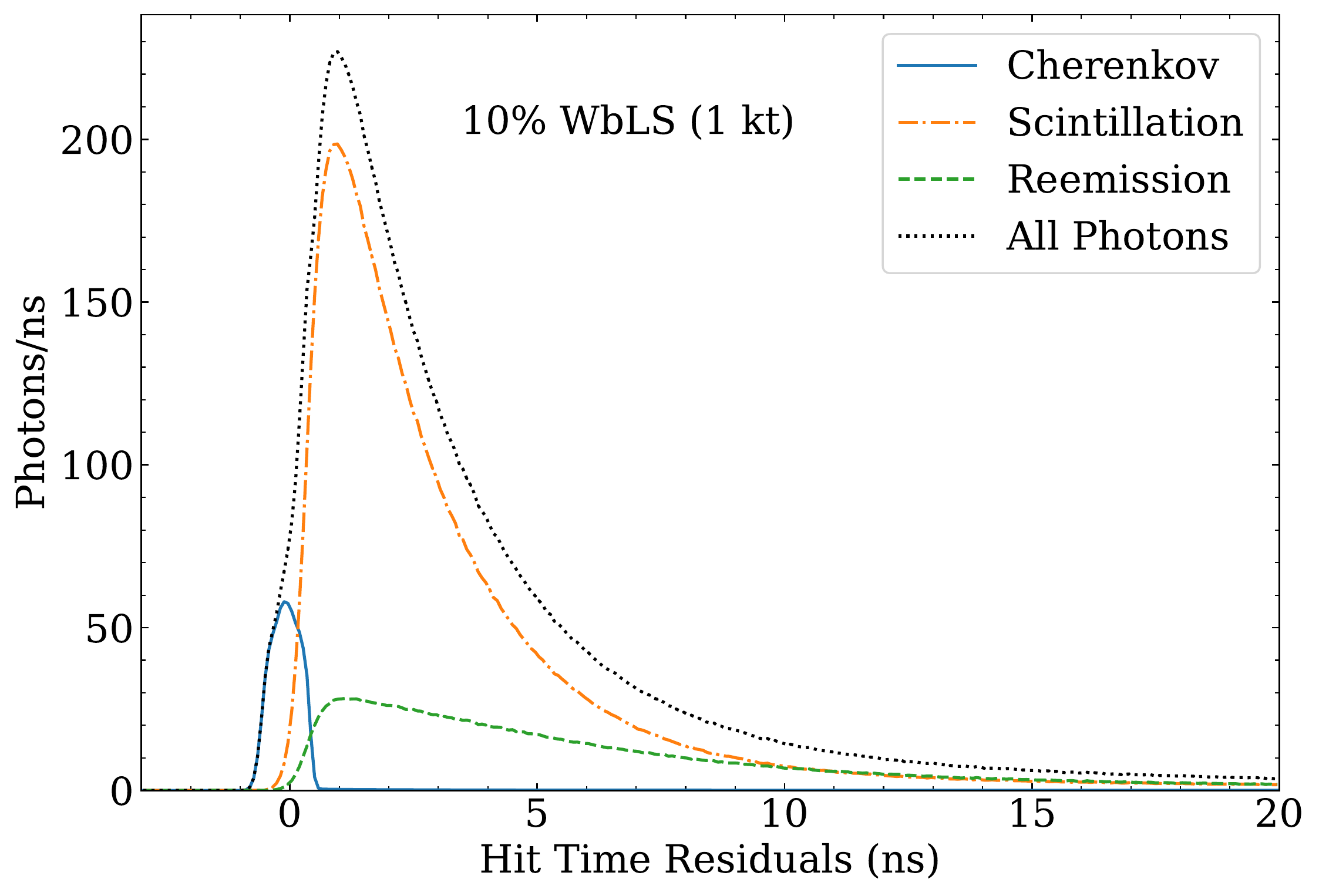}
\includegraphics[width=\columnwidth]{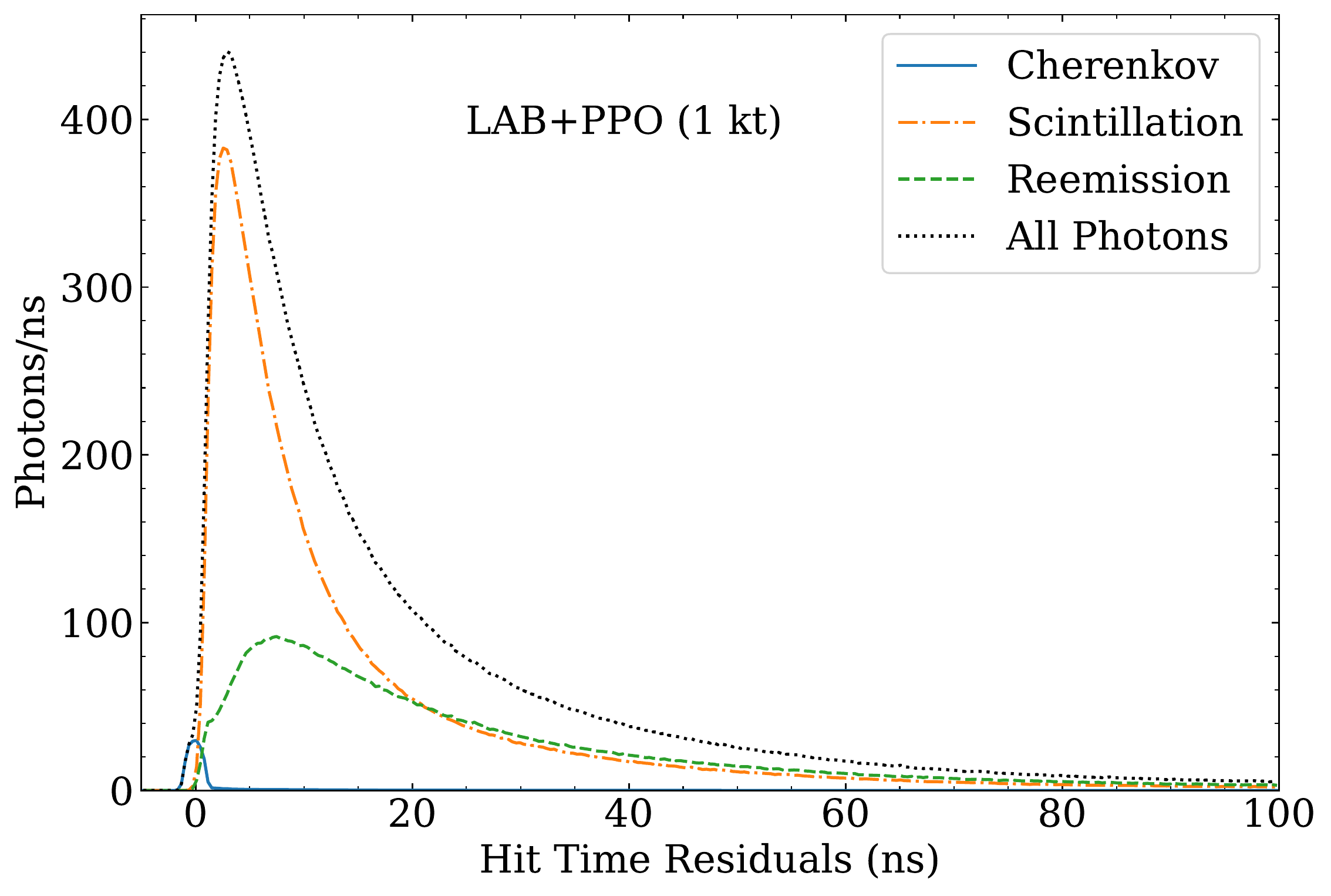} \\
\includegraphics[width=\columnwidth]{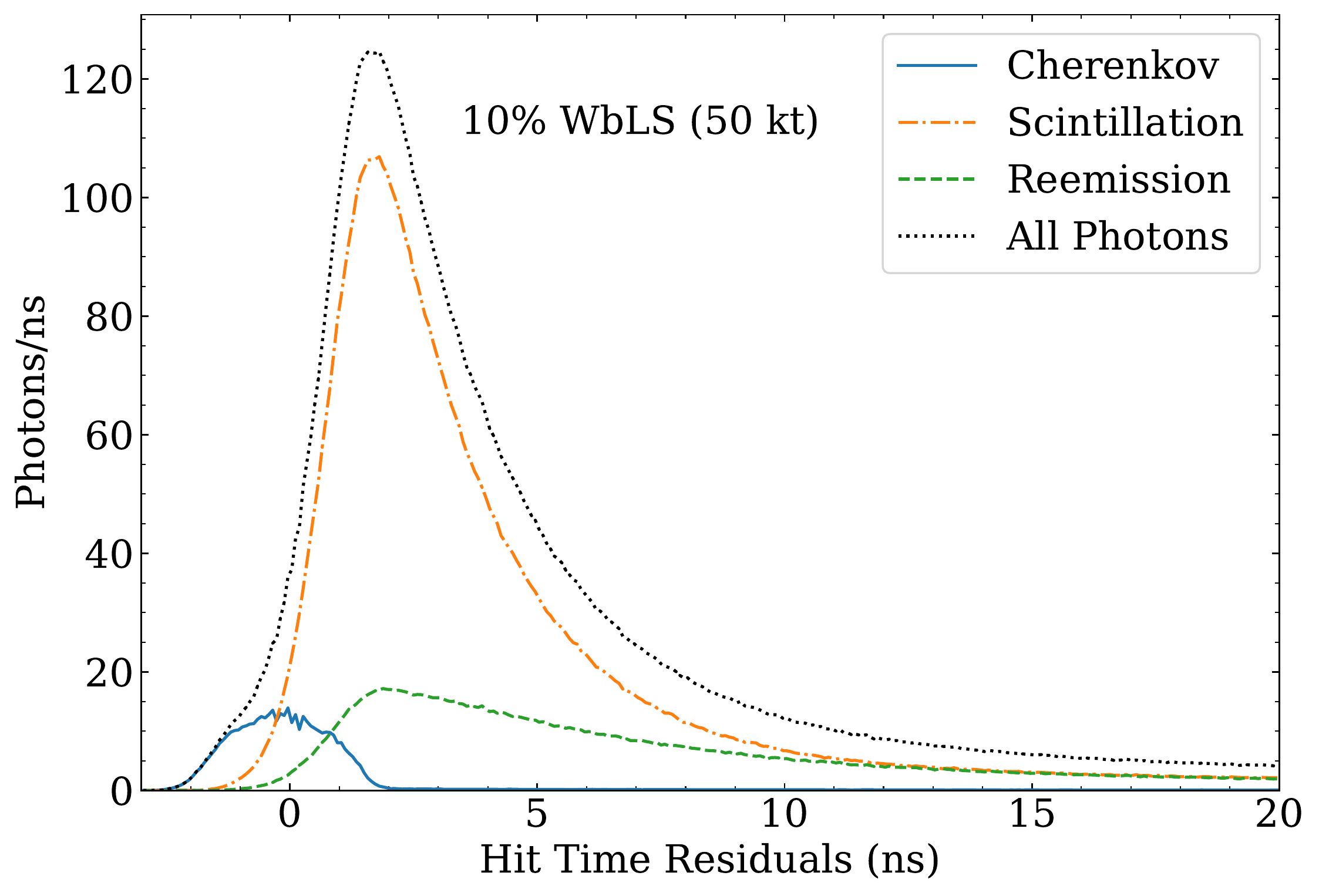}
\includegraphics[width=\columnwidth]{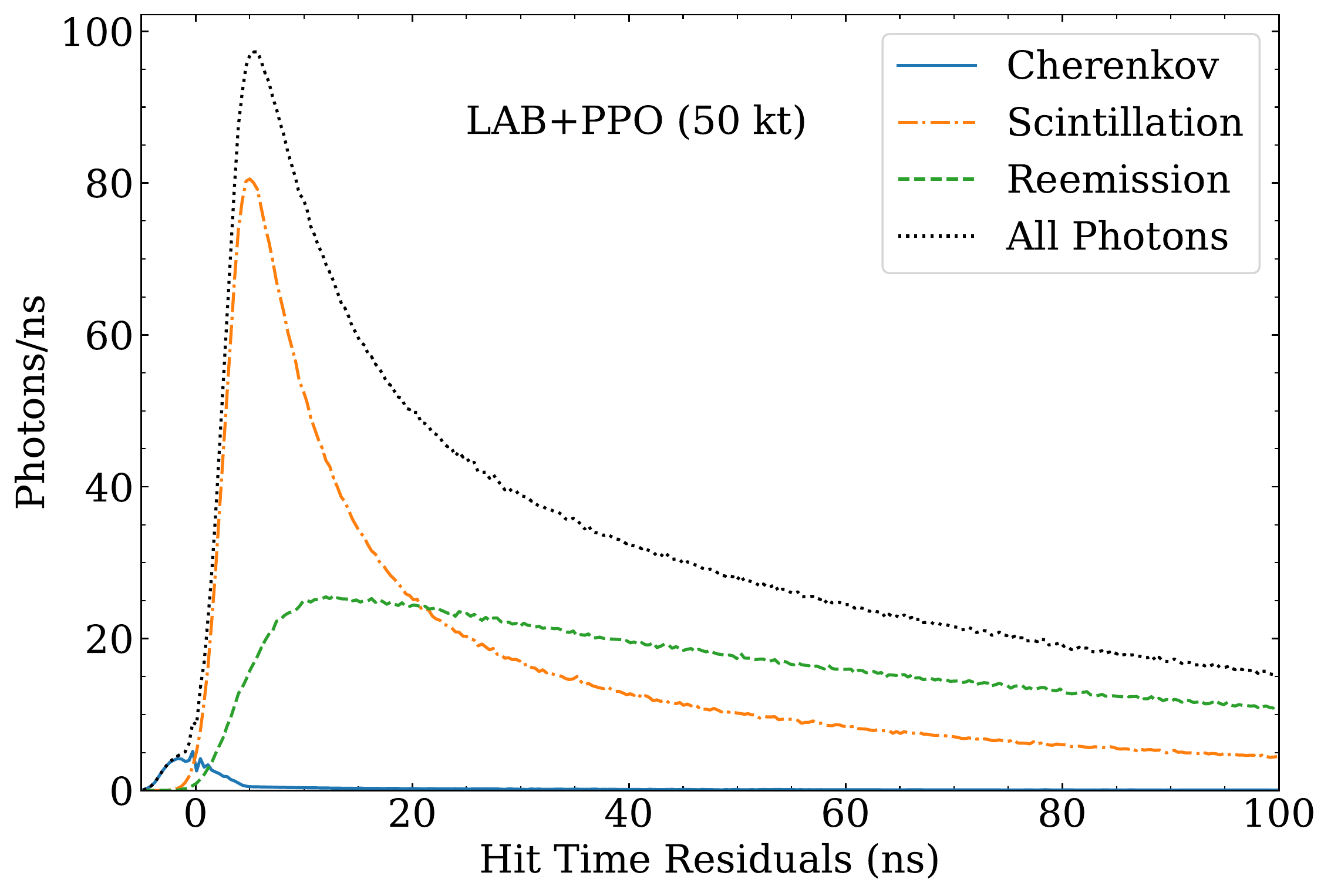} \\
\caption{\label{fig:hit-time-residuals} True hit time residual distributions for (left) 10\% WbLS and (right) pure LS in a (top) 1-kt and (bottom) 50-kt detector. This uses the same QE as the photon detector models, but with zero TTS. Fluctuations observed in these distributions are purely statistical.}
\end{figure*}

The reconstruction algorithm used here has the following steps, which are described in detail in the following sections:
\begin{enumerate}[Step 1:]
\item Position and time of the interaction vertex are reconstructed using all detected hits by maximizing the likelihood of the $t_{resid}$ distribution.
\item Direction is reconstructed using only prompt hits by placing a cut on the $t_{resid}$ distribution, obtained for the reconstructed value of position and time.
\item Finally, the total number of hits is recorded as an estimate of the energy of the event.
\end{enumerate}

The approach is inspired by 
vertex reconstruction algorithms used in the SNO 
experiment~\cite{sno_reconstruction}. The algorithm has been tested and 
demonstrated to achieve similar position and direction resolution to SNO for 
equivalent event types in a SNO-like detector---for example, for 5~MeV
electrons in a SNO-sized vessel, with TTS and photo-coverage set to relevant 
values (approximately 1.8~ns and 55\%, respectively) this algorithm 
achieves 27.4$^\circ$ angular resolution, compared to the SNO reported value 
of 27$^\circ$.

We note that this choice of reconstruction methodology is one that can be applied for the full spectrum of materials under consideration, from water to pure LS.  Significant work is ongoing in the community to develop reconstruction techniques specific to certain materials and certain detector configurations, or particular physics goals~\cite{Wonsak:2018uby,Wonsak_Dresden,Dunger:2019dfo,Aberle:2014,elagin,Jiang:2019cnb,Li:2018}.  Such methods would likely out-perform our approach when applied to the intended detector or physics goal, and it is highly likely that the results presented here can be further optimized by the incorporation of such algorithms.  As such, these results should be considered conservative.  Our intent is to apply a single algorithm across all materials to facilitate comparison between detector configurations.

\subsubsection{Position and time}
Reconstructing vertex position and time can be done by maximizing the likelihood of $t_{resid,\,i}$ for each hit $i$ in the event:
\begin{equation}
t_{resid,\,i} = \left(t_i - t\right) - \left|\vec{x}_i - \vec{x}\right| \frac{n}{c},
\end{equation}
where $(\vec{x}_i,t_i)$ are the position and time of a detected photon,  $(\vec{x},t)$ represents the fitted vertex position and time, and $\frac{c}{n}$ is the group velocity typical of a 400-nm photon.
This expression includes two important assumptions that are made to approximate a realistic detection scheme.
\begin{enumerate}
\item The travel time is calculated assuming a photon wavelength of 400 nm, since for a real detector the wavelength is typically not known.  Fig.~\ref{fig:qe} shows the expected spectra for both Cherenkov and scintillation light.
\item Each photon is assumed to travel in a straight line, as photon detectors are typically not aware of the actual path the photon traveled.
\end{enumerate}
A result of these assumptions is that dispersion in the material will broaden the $t_{resid}$ distribution, as the travel time will be overestimated (underestimated) for longer (shorter) wavelength photons.
Additionally, scattered or reemitted photons will appear later than their true emission time due to ignoring their true path.
An example of a $t_{resid}$ distribution using the true detection times, but with these approximations, is shown for the  10\% WbLS and pure LS material in \Cref{fig:hit-time-residuals} for the 1-kt and 50-kt detector geometries.
In plots shown in this paper, the $t_{resid}$ is arbitrarily shifted such that the average $t_{resid}$ of Cherenkov photons across many events is 0 ns.
The integral of these distributions is the number of detected photons per event on average, which highlights both the difficulty of identifying Cherenkov photons in pure scintillators, and their prompt placement in the $t_{resid}$ distribution.

For each material and detector configuration, a PDF for $t_{resid}$ of all photons is produced using truth information from a subset of the simulated events.
Reconstruction is then done by minimizing the sum of the negative logarithm of the likelihood for each hit with a two-staged approach: a Nelder-Mead~\cite{scipy} minimization algorithm with a randomly generated seed is used to explore the likelihood space and approximate the global minima, followed by a BFGS~\cite{scipy} minimization algorithm to find the true (local) minima using the minima from the previous step as the seed.
This method produces the best estimate of the true $t_{resid}$ distribution for each event, to be used in the direction fit.

For each event, the difference between the reconstructed position (time) and the true position (time) is taken.
The distributions of these differences for each material and detector configuration are fit to Gaussian distributions, and the sigma of these fits is taken as the resolution for the position and time reconstruction.
The position resolutions reported here are the quadrature sum of the widths in all three dimensions.

\subsubsection{Direction}
\begin{figure*}[]
\centering
\includegraphics[width=\columnwidth]{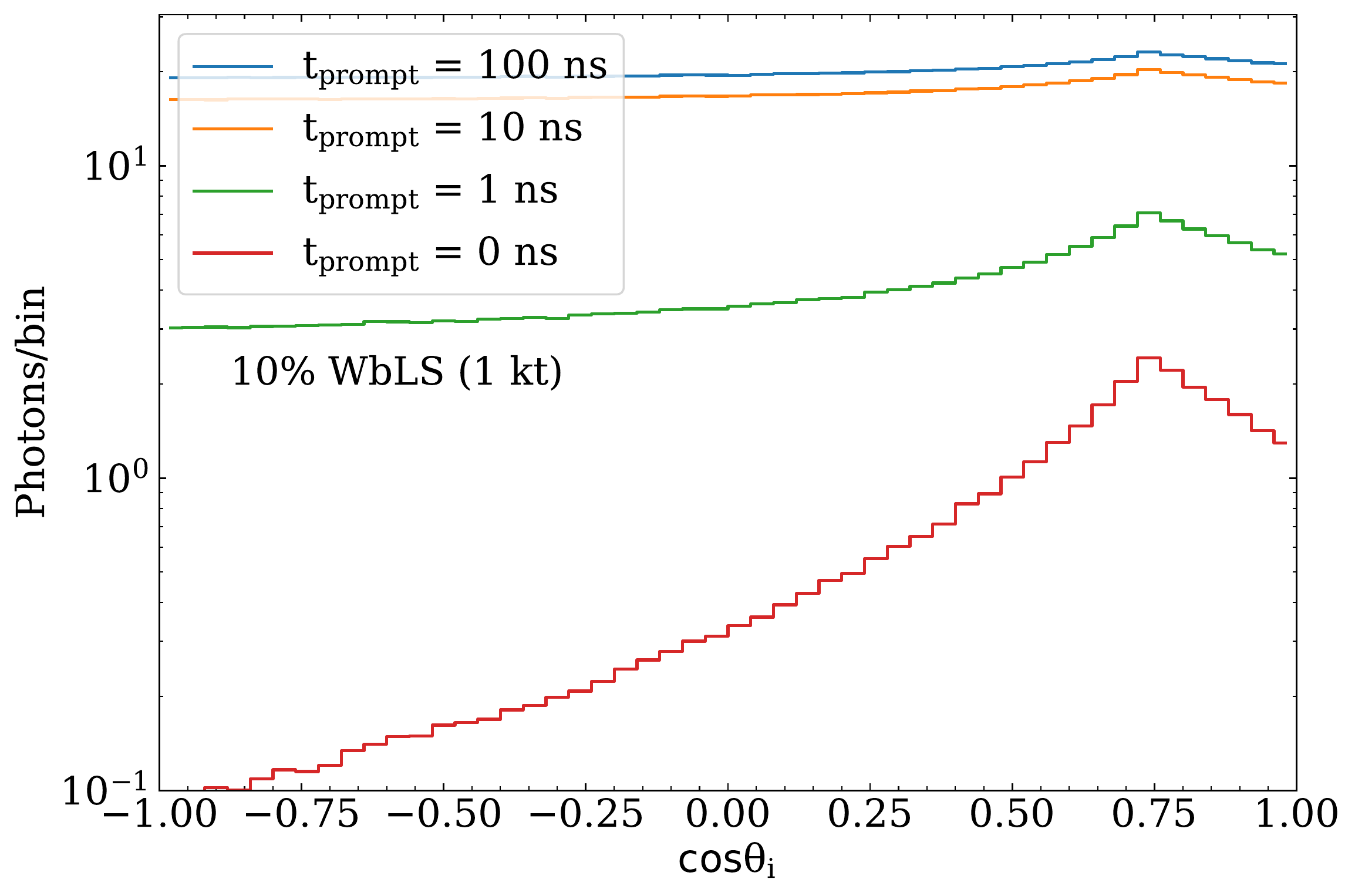} 
\includegraphics[width=\columnwidth]{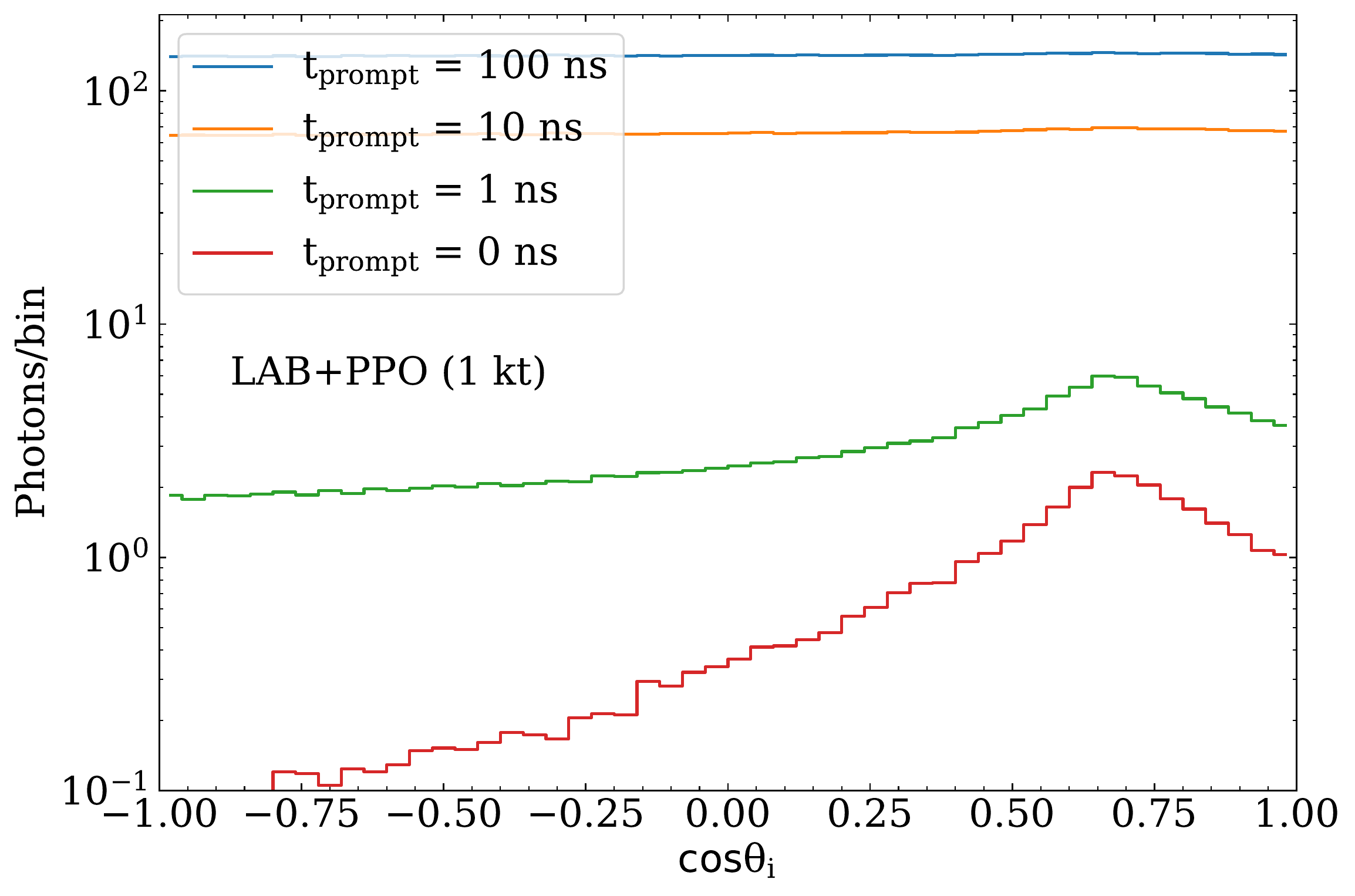} \\ 
\includegraphics[width=\columnwidth]{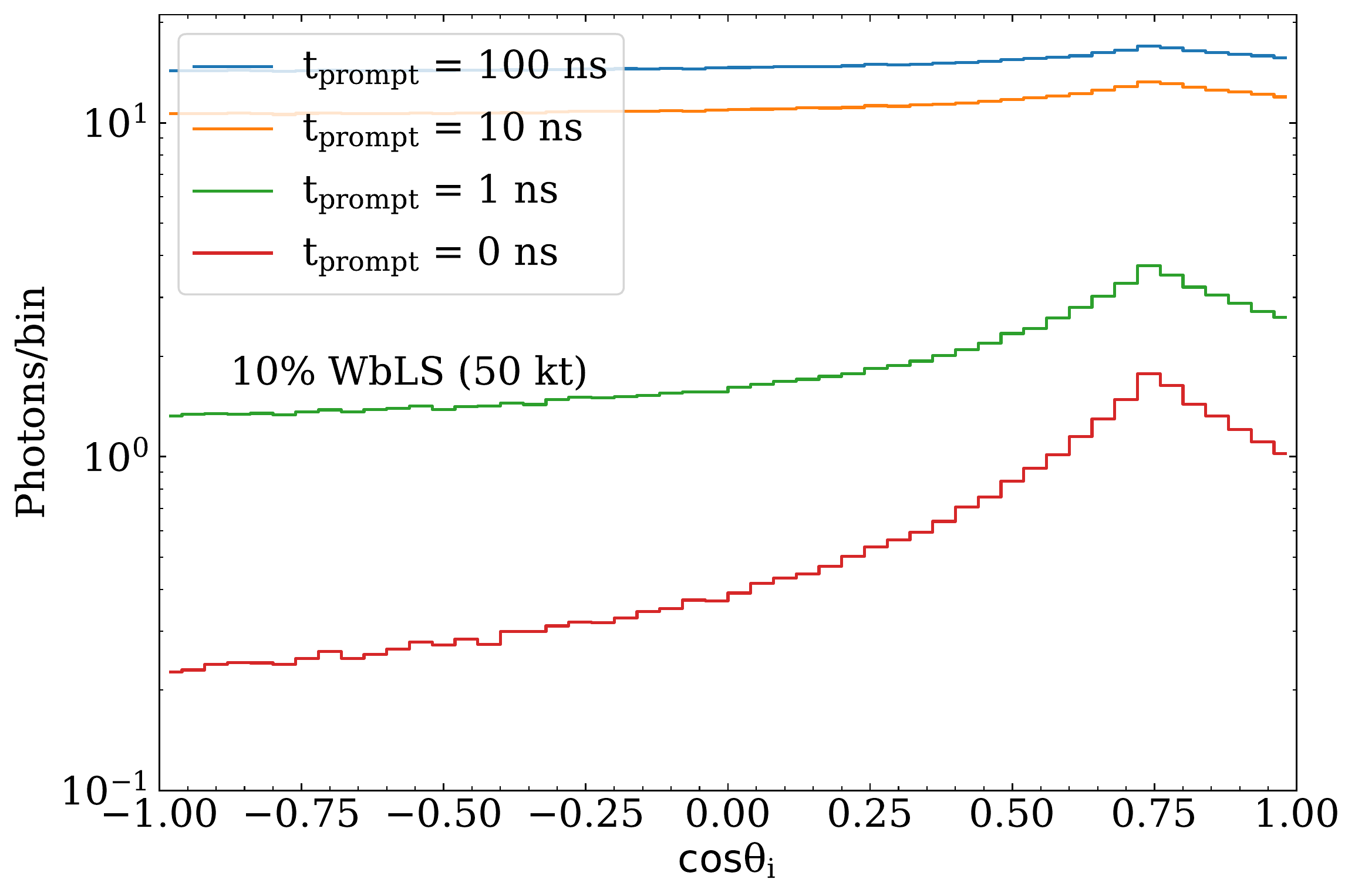} 
\includegraphics[width=\columnwidth]{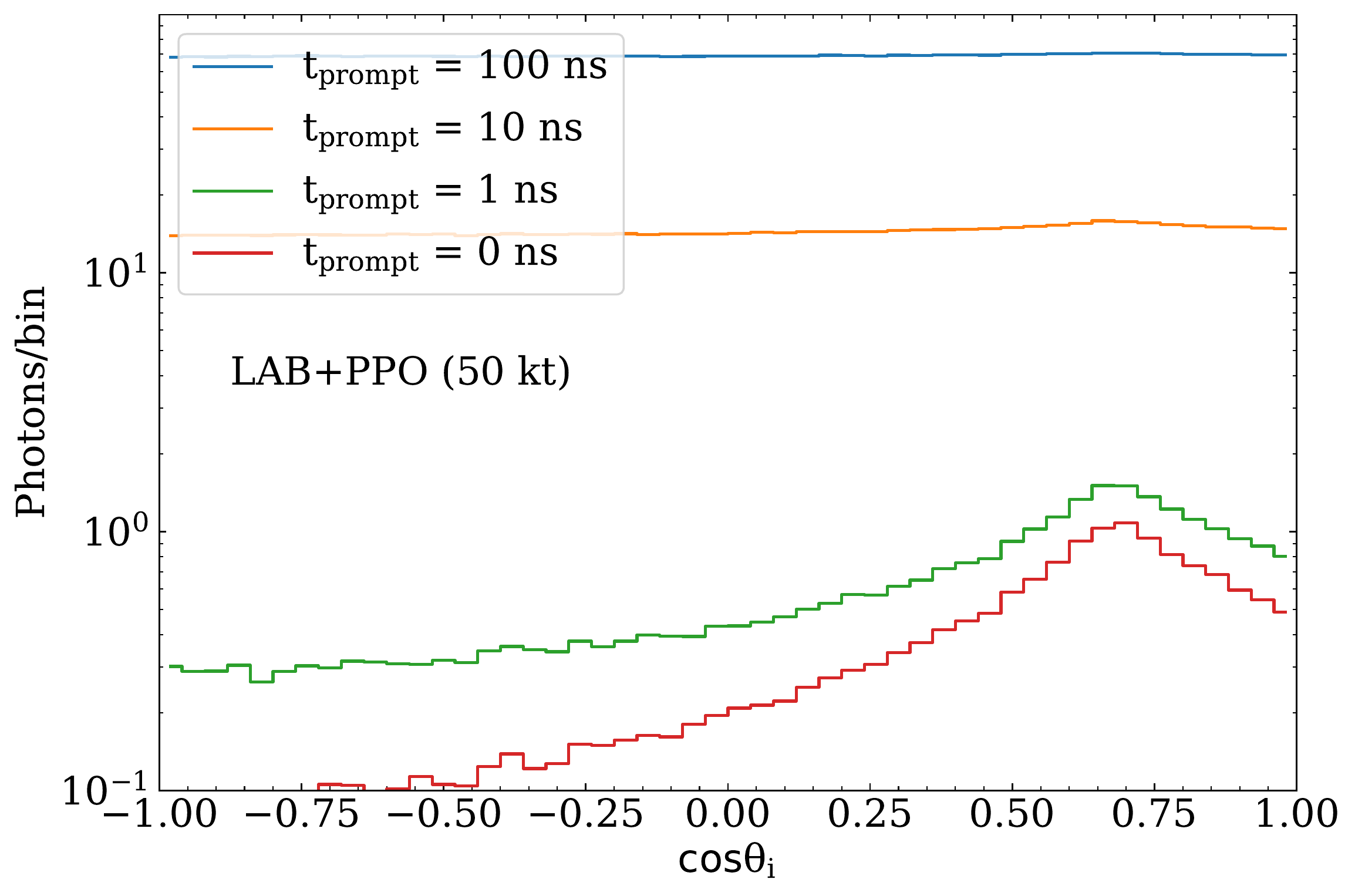} \\ 
\caption{\label{fig:costheta-distributions} True photon direction distributions for (left) 10\% WbLS and (right) pure LS in a (top) 1-kt and (bottom) 50-kt detector. These are shown for several $t_{prompt}$ cuts, highlighting how prompt cuts on the hit time residual distribution can reveal the directional Cherenkov photons, even in pure LS. Fluctuations observed in these distributions are purely statistical.}
\end{figure*}

As Cherenkov light is emitted at a fixed angle with respect to the particle's path, detected Cherenkov hits can be used to infer the event direction. 
A  method for doing this is by maximizing the likelihood of the cosine of the angle, $\theta_i$, between the vector from the reconstructed event position, $\vec{x}$, to each detected photon position, $\vec{x}_i$, and a hypothesized direction $\hat{d}$:
\begin{equation}
\cos{\theta_i} = \frac{\left(\vec{x}_i-\vec{x}\right) \cdot \hat{d}}{\left|\vec{x}_i-\vec{x}\right|}.
\end{equation}
For Cherenkov light, the PDF for this distribution is peaked at the Cherenkov emission angle, $\theta_c$, of the material.
Because non-Cherenkov photons do not carry directional information, they will appear flat in this distribution, and will degrade the performance of the fit.
It is beneficial, therefore, to restrict this likelihood maximization to only photons with $t_{resid} < t_{prompt}$ for some $t_{prompt}$, as this should maximize the number of Cherenkov photons relative to other photons.
Examples of the $\cos \theta_i$ distributions with various $t_{prompt}$ cuts is shown in \Cref{fig:costheta-distributions} for 10\% WbLS and pure LS.
These figures show that in the 10\% WbLS material, directional information is still visible even with large $t_{prompt}$ cuts, whereas this is not the case with pure LS, where the scintillation light greatly exceeds the Cherenkov light.
Here, the impact of dispersion is typically beneficial, as the broad spectrum of Cherenkov light compared to typical scintillation spectra results in long-wavelength Cherenkov photons appearing earlier in the $t_{resid}$ distribution compared to their true emission times.
We note that a photon detection scheme that can distinguish between long and short wavelength photons~\cite{dichroicon2} could further enhance the ability to identify Cherenkov photons.

PDFs for the $\cos{\theta_i}$ distribution are created using subsets of the simulated events for many $t_{prompt}$ values between -1 ns and 10 ns, and event reconstruction is done for each $t_{prompt}$ value for every event.
Reconstruction proceeds in the same way as the position-time minimizing the sum of the negative logarithms of the likelihood of each selected hit with a randomly seeded coarse Nelder-Mead~\cite{scipy} search, followed by a BFGS~\cite{scipy} method seeded with the result of Nelder-Mead to find the best minima. 
The value $\cos{\theta}$ is calculated for each reconstructed direction as $\hat{d} \cdot \hat{d}_{true}$, where $\hat{d}_{true}$ is the initial direction of the electron.
The $\cos{\theta}$ distribution from each simulated configuration and $t_{prompt}$ pair is integrated from $\cos{\theta} = 1$ until the $\cos{\theta}$ value that contains 68\% of events, and this value is defined as the angular resolution for that pair.
Finally, the angular resolution resulting from the $t_{prompt}$ with the best angular resolution for each configuration is taken as the angular resolution for that configuration.

\subsubsection{Energy}
The distribution of the total number of hits is fit to a Gaussian to determine the mean $\mu_N$ and standard deviation $\sigma_N$ of detected hits for each condition. 
The fractional energy resolution is reported as $\sigma_N / \mu_N$.

\section{\label{sec:impact} Performance of water-based liquid scintillator in a large-scale neutrino detector}

The materials described in \Cref{sec:methods} were simulated in the two detector geometries (1 kt and 50 kt) and four photodetector models (``PMT,'' ``FastPMT,'' ``FasterPMT,'' and ``LAPPD'') described in the same section. 
Between 10,000 and 100,000 events were simulated for each material, with fewer events for the pure LS due to the high photon counts (and accordingly slower simulation times). 
The following sections explore the true MC information provided by those simulations, as well as presenting the reconstruction results for all cases.


\subsection{Photon population statistics}

\begin{figure}[]
\centering
\includegraphics[width=\columnwidth]{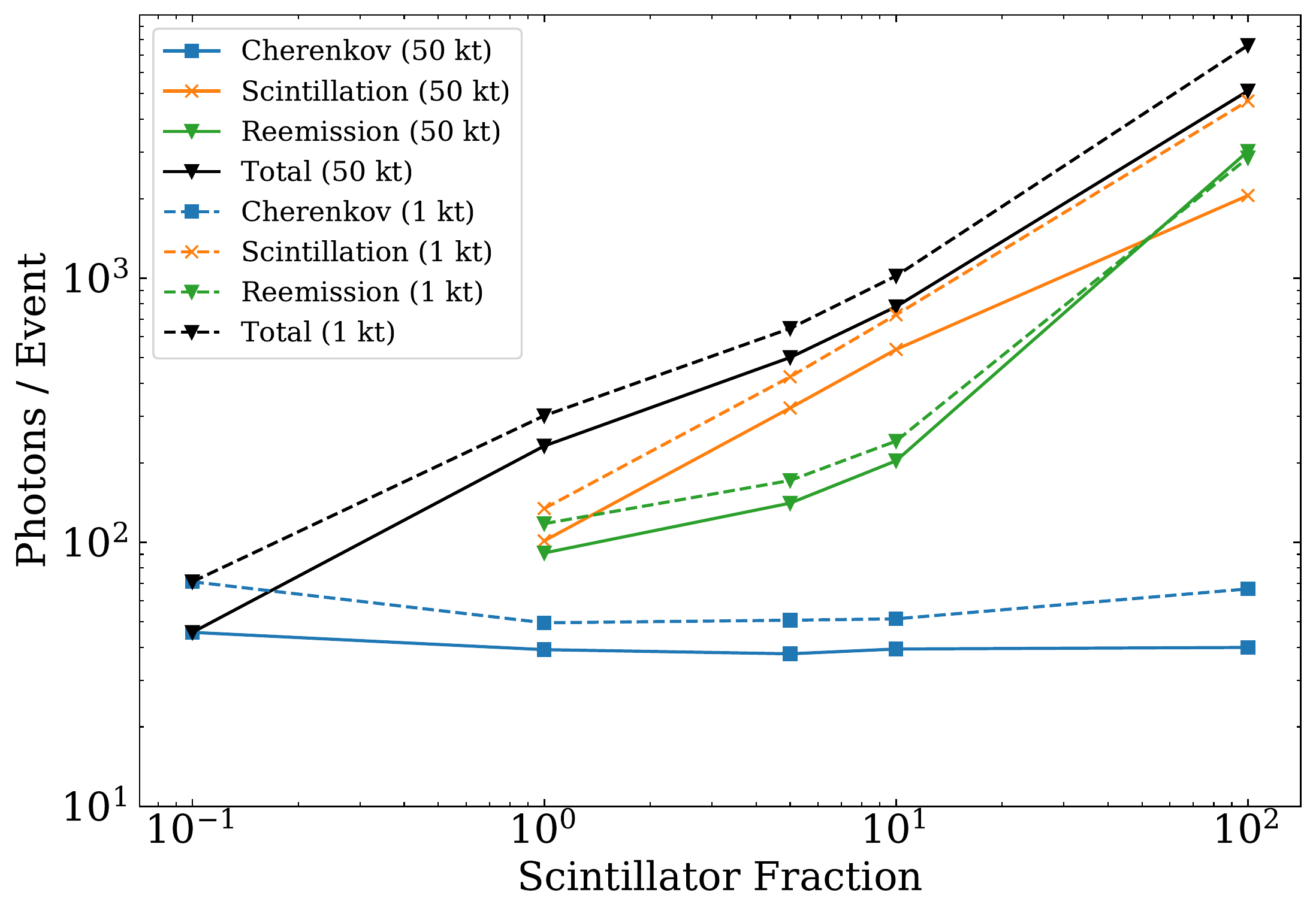}
\caption{\label{fig:photon-stats}The number of detected photons for 2.6-MeV electrons simulated at the center of two detector geometries (50-kt and 1-kt) differing in size. These photon counts are shown as a function of material scintillator fraction. Water is artificially plotted at $10^{-1}$ (due to log scale).}
\end{figure}

Roughly speaking, energy resolution is limited by the total number of detected photons, position and time resolution are limited by the number of direct photons (not absorbed and reemitted, scattered, or reflected), and direction resolution is limited by the number of Cherenkov photons and how visible they are within the brighter scintillation signal.
The total population of photons can be broken down into the following categories:
\begin{enumerate}
\item \textit{Cherenkov} photons, which were not absorbed and reemitted by the scintillator.
\item \textit{Scintillation} photons, which were not absorbed and reemitted by the scintillator.
\item \textit{Reemitted} photons, regardless of their origin.
\end{enumerate}
These populations are shown in \Cref{fig:photon-stats} for the materials and detector sizes considered here. 
Since each considered photon detector model has the same QE and coverage, the populations are the same in each case.

Higher scintillator fractions are very advantageous from an energy resolution perspective, having many more total photons. 
The same is true from the perspective of position and time resolution in a 1-kt detector.
For a larger 50-kt detector, the population of reemitted photons for pure LS is greater than the scintillation population, hinting that this condition is dominated by absorption and reemission, which can degrade vertex reconstruction, as reemitted photons are less correlated with the initial vertex. 
Despite the larger refractive index in pure LS, which implies a larger number of generated Cherenkov photons, the number of detected Cherenkov photons is highest in water in both detector sizes.
In the WbLS materials, the increase in refractive index is largely offset by the shorter attenuation lengths, resulting in a nearly flat trend for detected Cherenkov photons in the 50-kt detector.
For the 1-kt detector, the water and pure LS materials are slightly favored over WbLS in terms of detected Cherenkov photons.
The difference between the two detector sizes is primarily due to attenuation, where the larger size results in more Cherenkov photons being absorbed.
As the total number of detected Cherenkov photons is similar for materials within the same detector size, the relative amount of scintillation photons, and the extent to which they can be discriminated from Cherenkov photons with $t_{prompt}$ cuts, plays a large role in reconstruction performance. 

\subsection{In-ring photon counting}

Without applying reconstruction algorithms, one can inspect the truth information for the detected hits to understand their origins and time distributions.
Of interest here is how discernible the Cherenkov photons are, and how well they may be identified against a scintillation background.
Since Cherenkov photons are emitted at a particular angle $\theta_c$ with respect to the track of the charged particle, it is instructive to see how many hits are detected in the region $\theta_c \pm \delta$ (``in-ring'') with respect to the event direction.
Further, since Cherenkov photons are prompt with respect to scintillation photons, it is instructive to see these populations as a function of how early they arrive.
As in the reconstruction algorithm, this is defined in terms of the hit time residual, $t_{resid}$, where smaller $t_{resid}$ values are more prompt.

\begin{figure}[]
\centering
\includegraphics[width=\columnwidth]{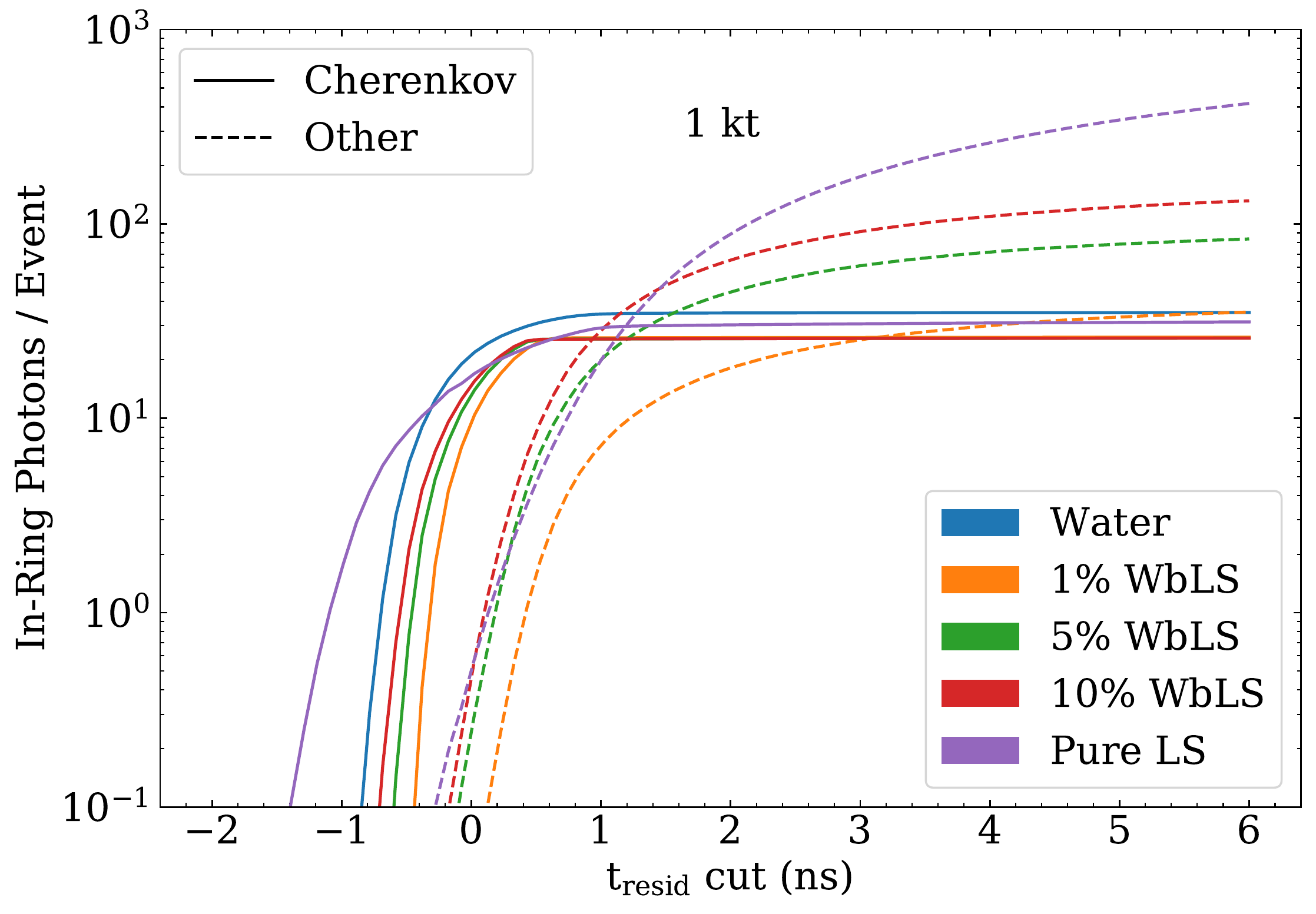}
\includegraphics[width=\columnwidth]{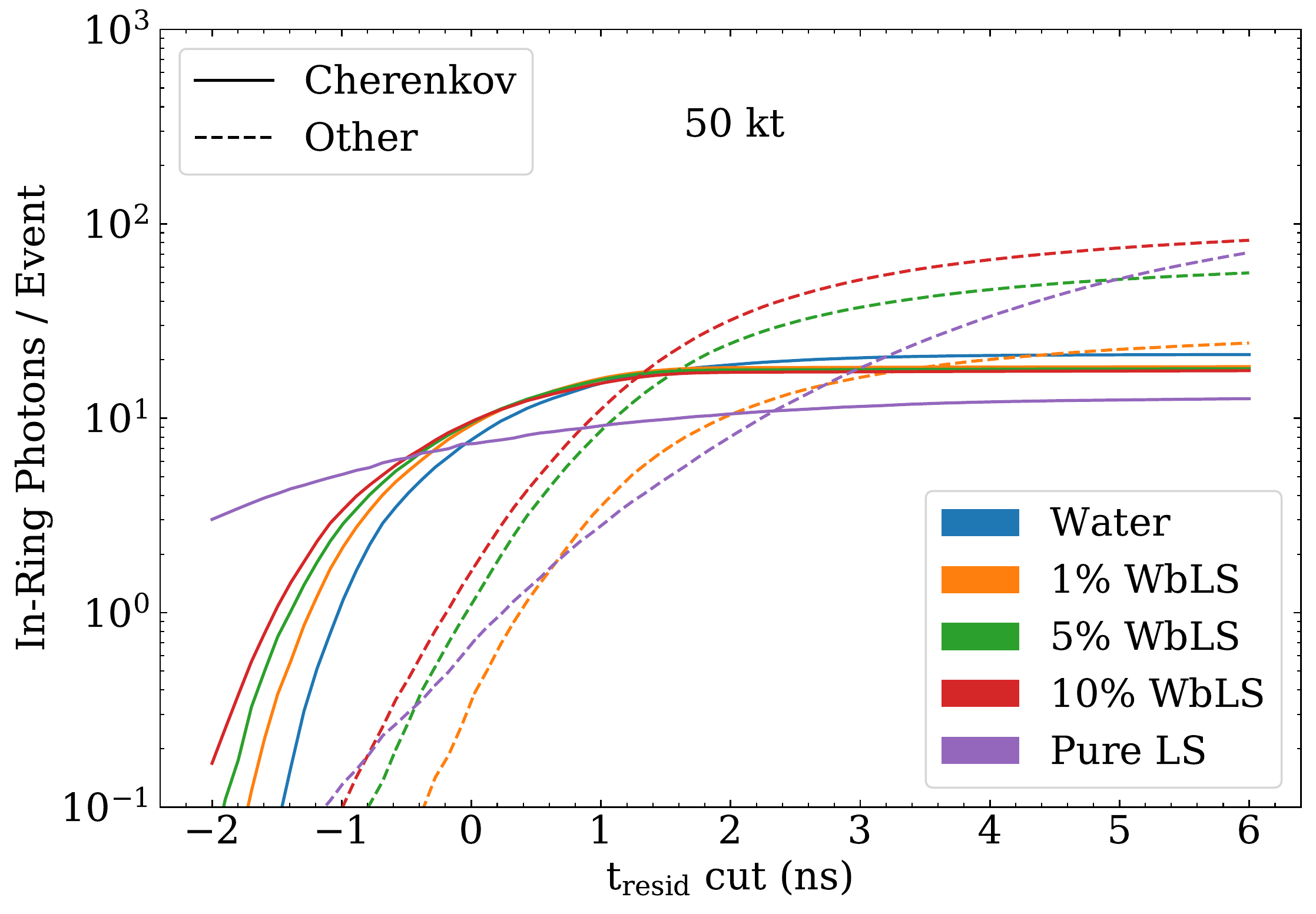}
\caption{\label{fig:in-ring-truth}The number of ``in-ring'' (see text) photons per event determined using truth information from 2.6 MeV electrons simulated at the center of two detector geometries (top) 1 kt and (bottom) 50 kt. The number of photons is shown as a function of $t_{prompt}$ cut, selecting for prompt photons. Cherenkov photons are shown in solid lines, with all other photons shown with dashed lines. The colored legend applies to both Cherenkov and other photons.}
\end{figure}

\Cref{fig:in-ring-truth} shows the number of Cherenkov and other (scintillation and re-emitted) photons for photons with $\cos \theta$ satisfying $\theta_c \pm 15^\circ$ using true detected times (TTS = 0) and true origins, but including the effect of photodetector coverage and QE, as a function of a $t_{prompt}$ cut on $t_{resid}$.
Of particular note is that there are more ``in-ring'' Cherenkov photons than other photons for sufficiently prompt $t_{prompt}$ cuts for all materials using truth information. 

\begin{figure}[]
\centering
\includegraphics[width=\columnwidth]{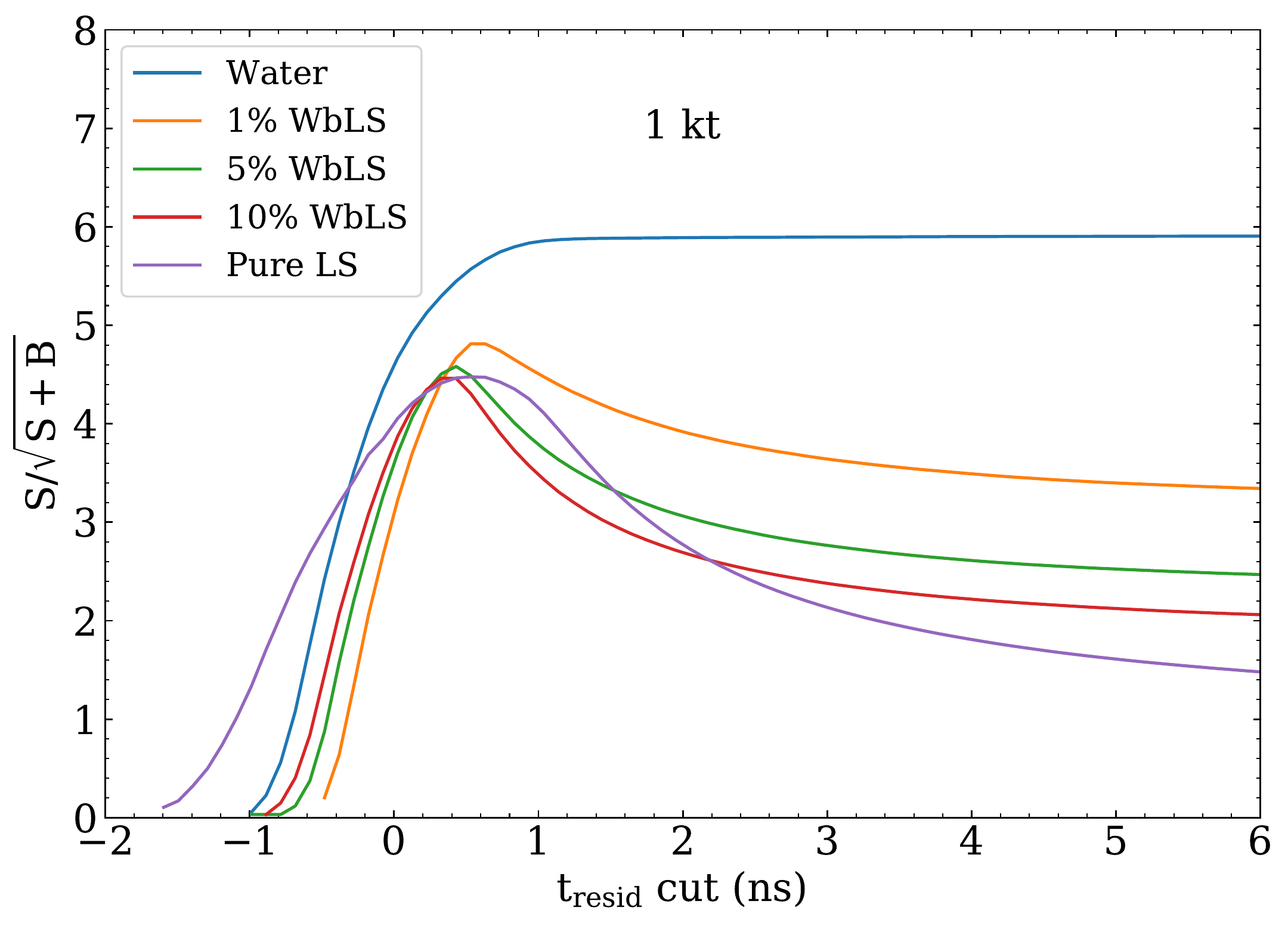}
\includegraphics[width=\columnwidth]{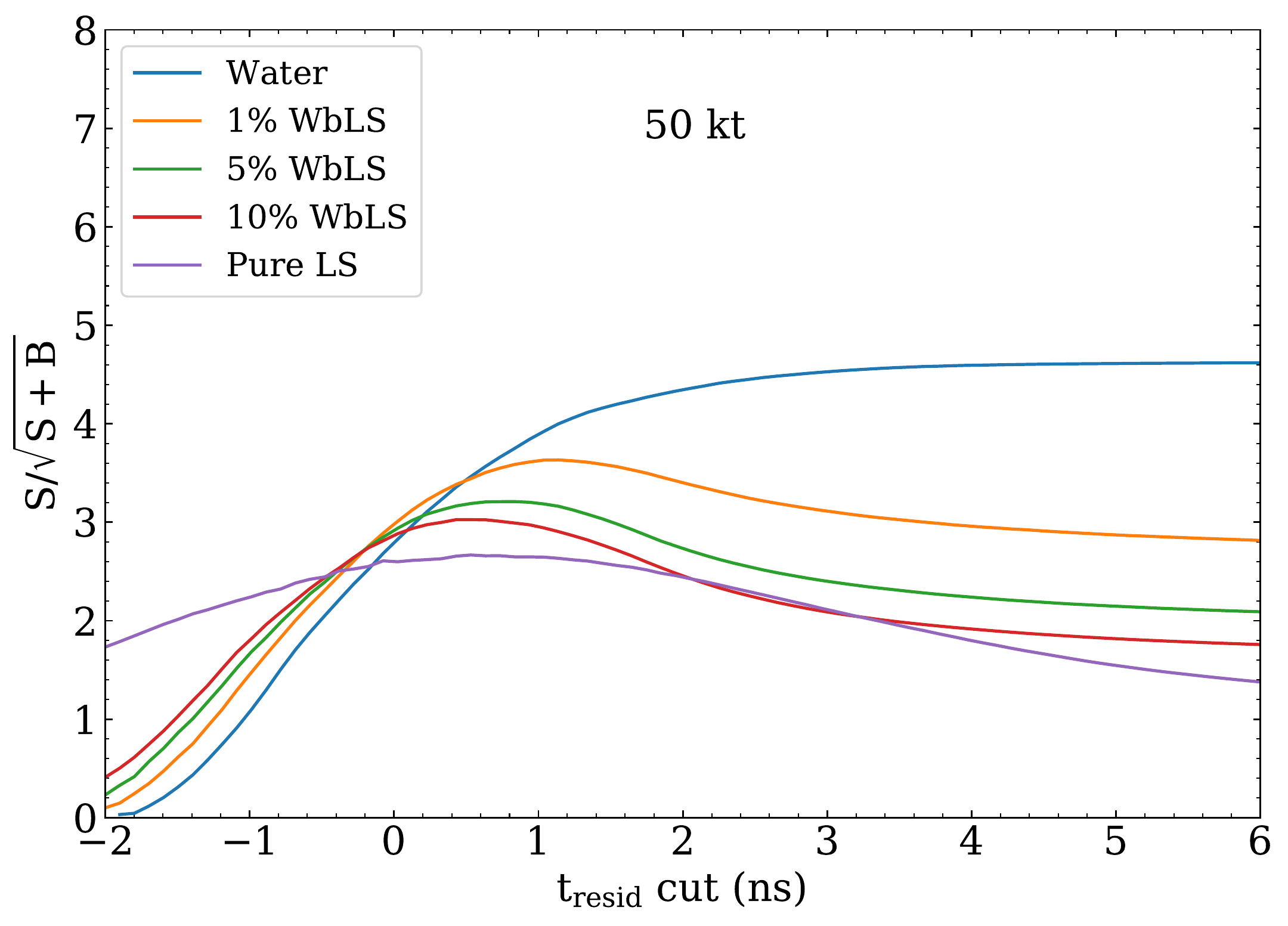}
\caption{\label{fig:in-ring-significance2}With $S$ defined as Cherenkov photons and $B$ defined as other photons, these figures plot $S/\sqrt{S+B}$, or the significance of the population of ``in-ring'' Cherenkov photons, for the data shown in \Cref{fig:in-ring-truth}, with the two detector geometries (top) 1 kt and (bottom) 50 kt. As this metric is only based on photon statistics and not reconstruction performance, it is used to inform, but not choose, the ideal $t_{prompt}$ cut (see \Cref{app:cuts}).}
\end{figure}

With the number of in-ring Cherenkov photons defined as $S$ and the number of in-ring other-photons defined as $B$, a single metric, $S/\sqrt{(S+B)}$, for the significance of the Cherenkov photons as a function of a $t_{prompt}$ cut is shown in \Cref{fig:in-ring-significance2}.
The larger this significance, the easier it should be to identify the Cherenkov topology on top of the isotropic scintillation background.
The higher significance at earliest times in the pure LS material is primarily due to the larger impact of dispersion in this material relative to WbLS or water.
Dispersion separates the narrow scintillation spectrum from the longer-wavelength portion of the broad-spectrum Cherenkov photons in large detectors, pushing the long-wavelength Cherenkov earlier, and the short-wavelength scintillation (and short-wavelength Cherenkov) later.
This results in better time separation between the earliest Cherenkov photons and the earliest scintillation photons when comparing pure LS to WbLS.
 
Also of note here is the similar amounts of prompt scintillation light in the WbLS and pure LS materials, despite having very different amounts of total scintillation light.
This is particularly clear in the 1-kt detector, where 5\%, 10\% and pure LS are very similar, while in the 50-kt detector those WbLS materials show more scintillation than pure LS at early times.
Two effects are at play here: differing amounts of dispersion due to differences in the refractive index, and also differences in the time profiles of the scintillation light in the different materials.
The effects of dispersion serve to delay the predominantly blue scintillation relative to the longer-wavelength Cherenkov light, and this occurs to a greater degree in the pure LS than in WbLS, due to the higher refractive index.
Further, the scintillation time profile of WbLS materials is faster than pure LS, as can be seen in the measurements from~\cite{chess_wbls}.
The combined effect is that there are similar amounts of prompt Cherenkov and prompt scintillation photons in the WbLS and pure LS materials, resulting in similar Cherenkov-significance in these materials.
As the scintillation light tends to come slightly later and is dimmest in 1\% WbLS, the greatest significance of Cherenkov detection in scintillating materials is achieved in that material, which also has the least stringent requirement on $t_{prompt}$ cut for peak performance. 
Both the 5\% and 10\% WbLS materials require an earlier $t_{prompt}$ cut than pure LS for peak performance, however more prompt cuts do result in slightly better Cherenkov significance than achieved in pure LS.

\subsection{\label{sec:recon_results}Reconstruction results}

\begin{figure}[]
\centering
\includegraphics[width=\columnwidth]{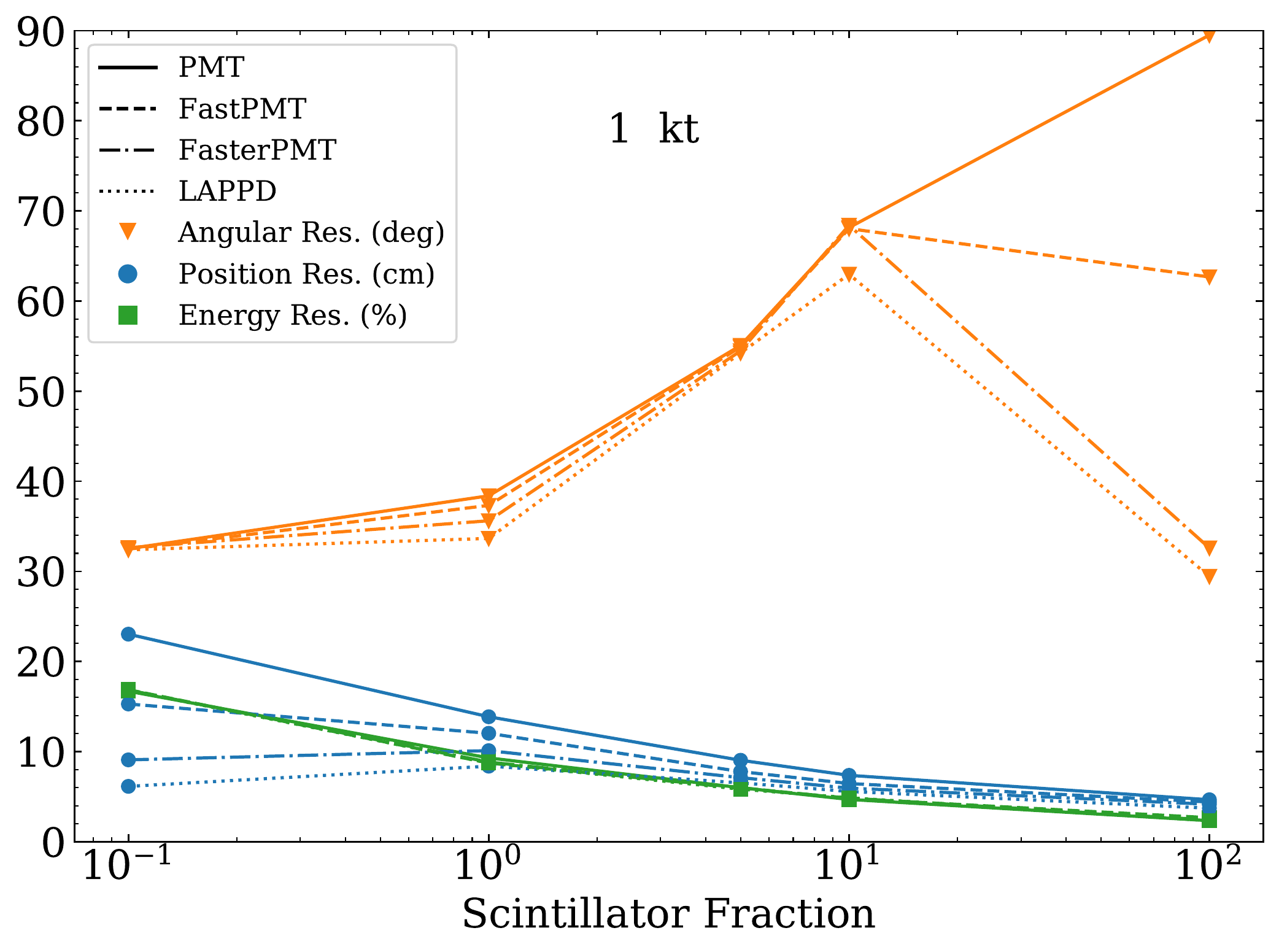}
\includegraphics[width=\columnwidth]{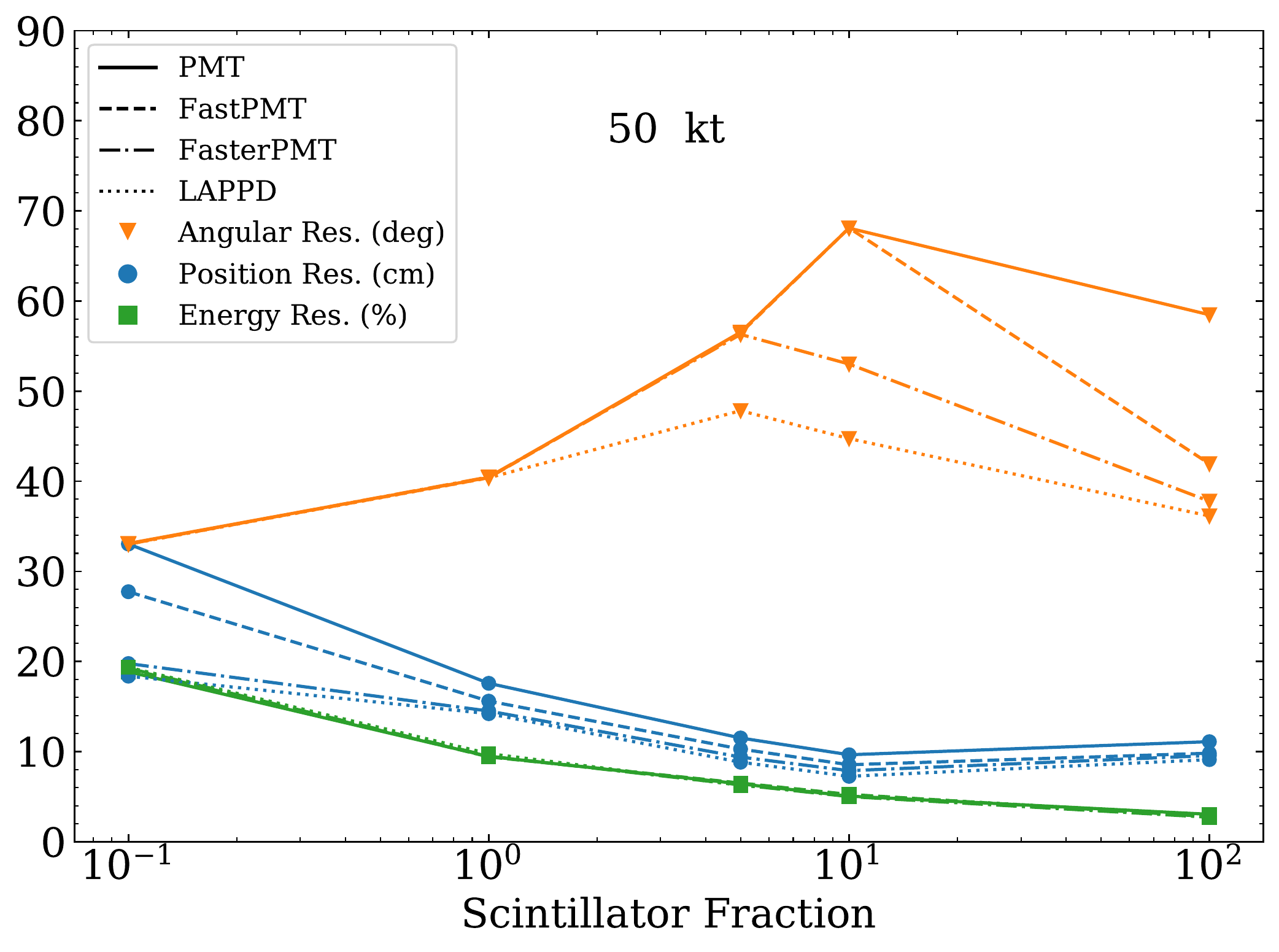}
\caption{\label{fig:reconstruction-metrics}Reconstruction resolutions of 2.6 MeV electrons simulated at the center of two detector geometries (top) 1-kt and (bottom) 50-kt, differing in size, and four photon detector models (``PMT,'' ``FastPMT,'' ``FasterPMT,'' and ``LAPPD''), differing in TTS. These resolutions are shown as a function of scintillator fraction. Water is artificially plotted at $10^{-1}$ (due to log scale). Angular resolution is shown for the best $t_{prompt}$ cut (see \Cref{app:cuts}). See legend for units.}
\end{figure}

\begin{figure}[]
\centering
\includegraphics[width=\columnwidth]{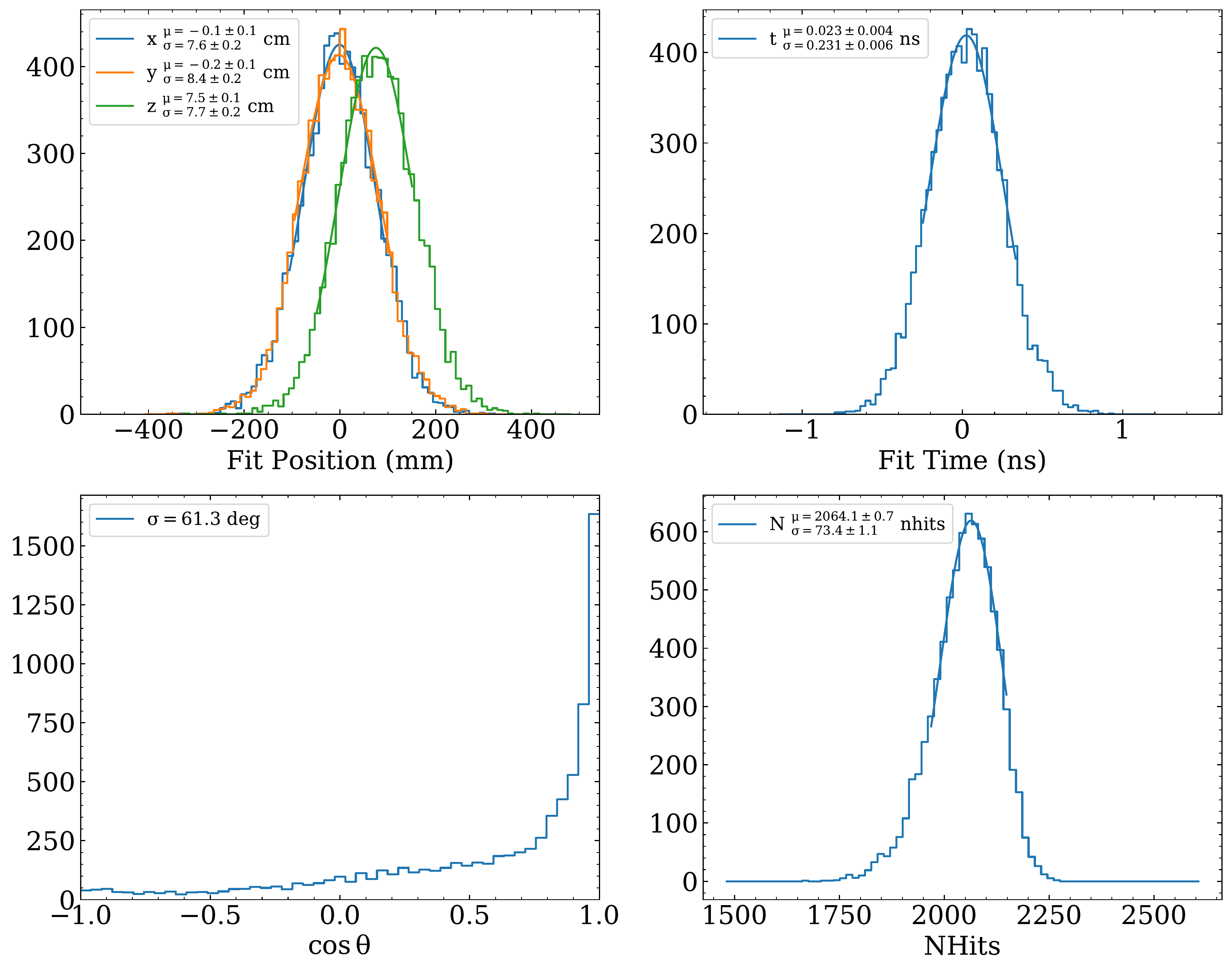}
\caption{\label{fig:reconstruction-detail} The upper left panel shows the position fit residuals in three dimensions, where $Z$ is always aligned with the initial event direction. The top right panel shows the fitted time residuals. The $\cos{\theta}$ fitted event direction distribution is in the bottom left, with the bottom right being the total number of detected photons, from which the energy resolution is calculated. This is shown for the pure LS material in the 1 kt detector geometry using the ``PMT'' photon detector model and a $1.0$~ns $t_{prompt}$ cut for direction reconstruction.}
\end{figure}

Inspecting the truth information provides a detailed understanding of the information available.
However, to truly evaluate these materials, it is necessary to apply reconstruction algorithms and evaluate the impact on position, time, and direction reconstruction.
This is done using the reconstruction algorithm described in \Cref{sec:methods} and the results are shown in \Cref{fig:reconstruction-metrics}.
An example view of the fit residuals for pure LS with a $1.0$~ns $t_{prompt}$ cut, showing the Gaussian fits to those residuals, can be found in \Cref{fig:reconstruction-detail}.
These results are a function both of material properties and the reconstruction algorithm used, and therefore should not be taken as the best possible resolutions achievable when using these materials.

In general, the scintillator materials outperformed water in the metric of position and time resolution due to the much larger number of photons detected from scintillation light.
The 1-kt detector typically demonstrates smaller residuals in position and time compared to the 50-kt detector, as the impact of dispersion and scattering, which broaden the $t_{resid}$ distribution, are greater in the larger geometry.
In particular, the better transparency of WbLS compared to pure LS is evident in the relatively poorer position resolution seen with pure LS when compared to 10\% WbLS in the 50-kt detector.
Position and time resolutions unsurprisingly improve with the reduction in TTS from the PMT model to the LAPPD model.

For direction reconstruction, the water material acts as an excellent baseline with best resolution, having only Cherenkov hits and excellent transparency.
The additional scintillation light from the WbLS materials degrades this resolution by approximately a factor of two in the 1-kt detector, and by less than 1.5 in the 50-kt detector for 10\% WbLS when using fast photodetectors like LAPPDs.
For pure LS, dispersion (especially in the 50-kt detector) and the relatively slower time profile results in enhanced $t_{resid}$ separation between Cherenkov and scintillation photons, enabling comparable or better angular resolution than the WbLS materials.
Notably, the LAPPD model has sufficient time resolution to easily identify a pure population of prompt Cherenkov photons in pure LS resulting from dispersion, allowing direction reconstruction comparable to water.
This is not seen with the PMT model, which lacks the time resolution to resolve this population.
This indicates that the dispersion of a pure scintillator is a beneficial quality for direction reconstruction, and that the faster timing profiles of the WbLS materials relative to pure LS may be a hindrance to accurate direction reconstruction.
The former point may be difficult to address in WbLS, given that the refractive index is very close to that of water and it is hardly tunable without significantly altering the material.
However, the time profiles of liquid scintillators can be adjusted~\cite{bnl_slow_scintillator,steve_slow_scintillator}, and this is explored in the following section.

\section{\label{sec:timing-impact} Impact of scintillation time profile in a large-scale neutrino detector}

As demonstrated in~\cite{chess_wbls}, the WbLS time profiles are faster than that of pure LS. It is useful to understand to what extent this difference impacts the performance of WbLS and pure LS. 
This can be studied by artificially adjusting the profiles of pure LS in simulation to match those of WbLS, and the reverse.
This also serves as first-order approximation of slow scintillators, and generally how adjusting the scintillation time profile impacts reconstruction.
What this approach does not take into account are the more complicated optics involved in the absorption and reemission of a secondary fluor, which would be present in slow scintillators~\cite{bnl_slow_scintillator,steve_slow_scintillator}.
Besides impacting the time profile, real fluors may have many other effects, such as reemission of photons at different wavelengths than the primary scintillation light, which could modify the impact of attenuation, dispersion, the matching of the spectra to the photodetector QE, among other things. 
However, this approach does explore to what extent the faster time profiles of WbLS impact its performance compared to pure LS, and what may be gained by exploring slower WbLS materials, perhaps by reducing the concentration of PPO~\cite{bnl_slow_scintillator}.

\begin{figure}[]
\centering
\includegraphics[width=\columnwidth]{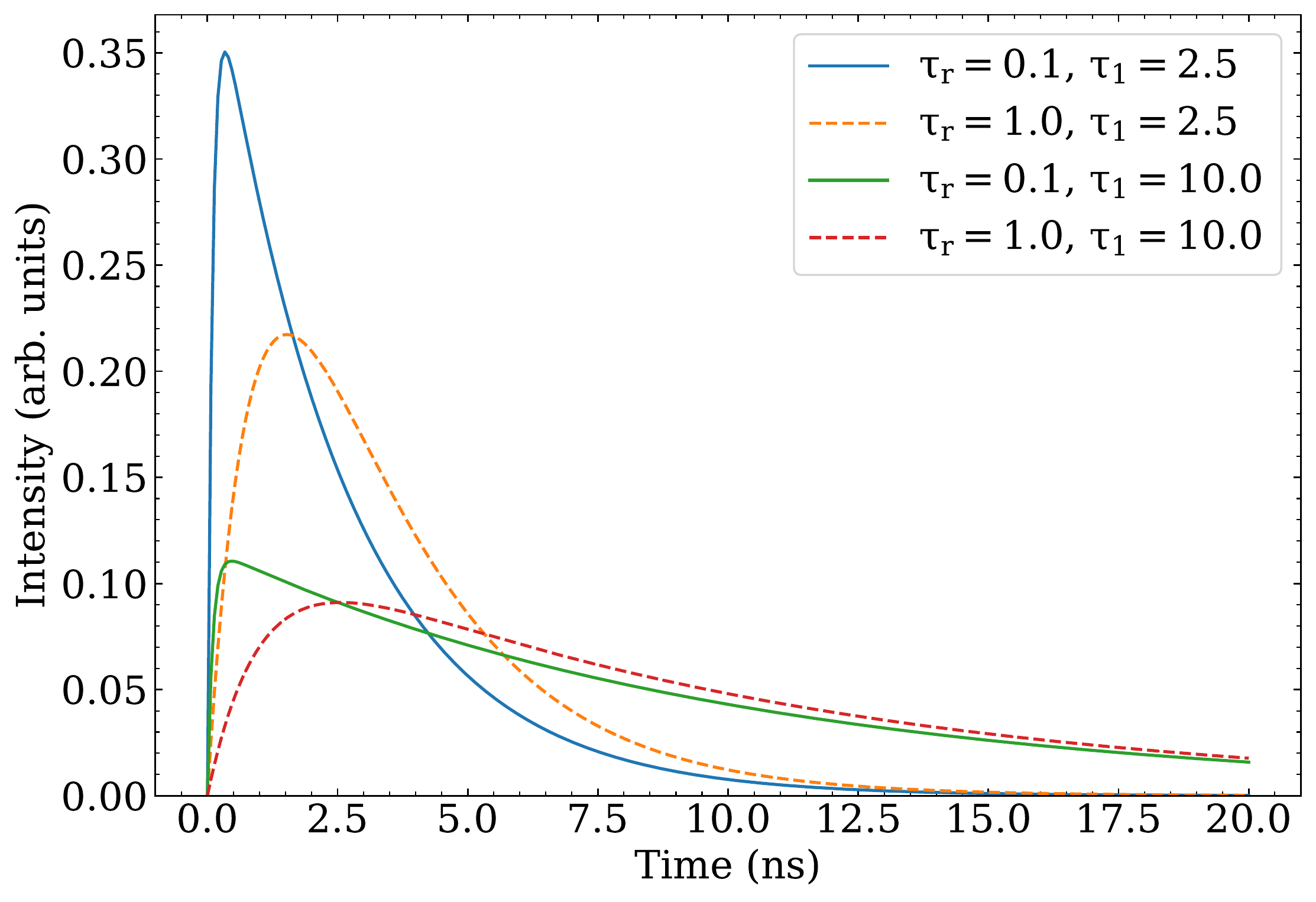}
\caption{\label{fig:tprofile} Example time profiles of the form \Cref{eq:tprofile}. The profiles are shown normalized to unit area, and cover the range of parameters used in the rise and decay time study.}
\end{figure}

Two properties are explored here: the rise time of the profile, $\tau_r$, and a single decay constant, $\tau_1$, using the form:
\begin{equation}
\label{eq:tprofile}
p(t) = \dfrac{1}{N}(1 - e^{t/\tau_1-t/\tau_r} ) e^{-t/\tau_1},
\end{equation}
where $N$ is a normalization constant.
Qualitatively, the decay time changes the amount of time over which the scintillation light is spread, with a larger decay time resulting in a broader emission profile.
The rise time, on the other hand, tends to delay earliest scintillation light without strongly impacting the overall width of the emission profile.
\Cref{fig:tprofile} visually shows the impact of changing these two parameters. 

Both the pure LS and 10\% WbLS materials have their time profiles adjusted, and reconstruction metrics are shown using the methodology described in \Cref{sec:methods}.
We consider both a scan of the decay constant for two chosen rise times, and a scan of the rise time for two chosen decay times.
In all cases, all other properties of the materials (light yield, refractive index, absorption and scattering, emission) are kept constant at the values presented in \Cref{sec:model}.
This allows us to decouple the effect of the time profile from other properties of the scintillator, which may be useful input for guiding future material development.

\subsection{Decay time}

The decay constant is scanned from 2.5 ns (typical of current WbLS) to 10 ns (typical of slow scintillators~\cite{bnl_slow_scintillator,steve_slow_scintillator}), and the simulation and reconstruction methods described in \Cref{sec:methods} are used for each combination.  This scan is repeated for two choices of rise time:
a fast rise time of 100~ps is used, characteristic of the WbLS cocktails explored in this paper, and a slow rise time of 1~ns, more representative of pure LS. 

\begin{figure*}[]
\centering
\includegraphics[width=\columnwidth]{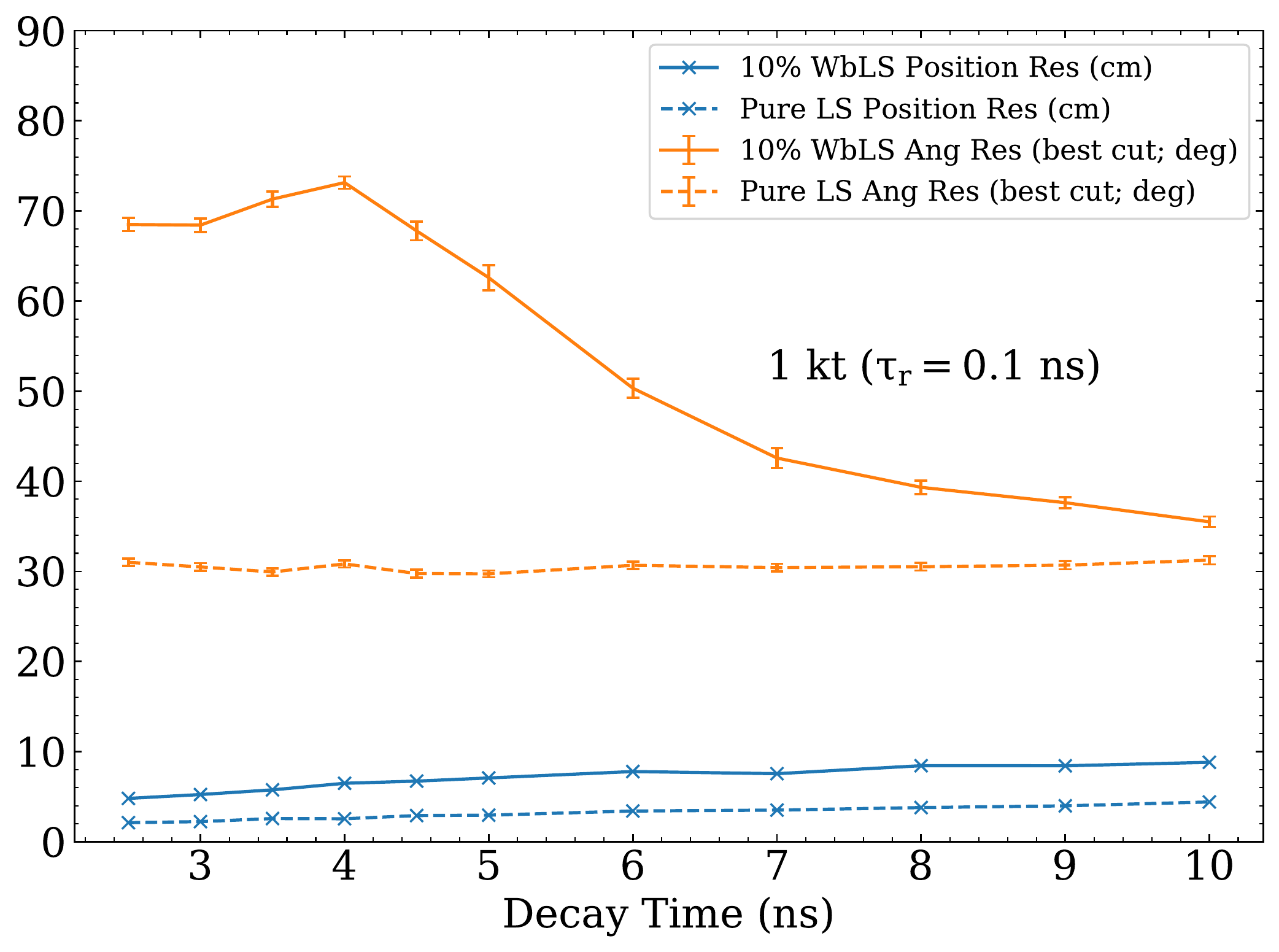}
\includegraphics[width=\columnwidth]{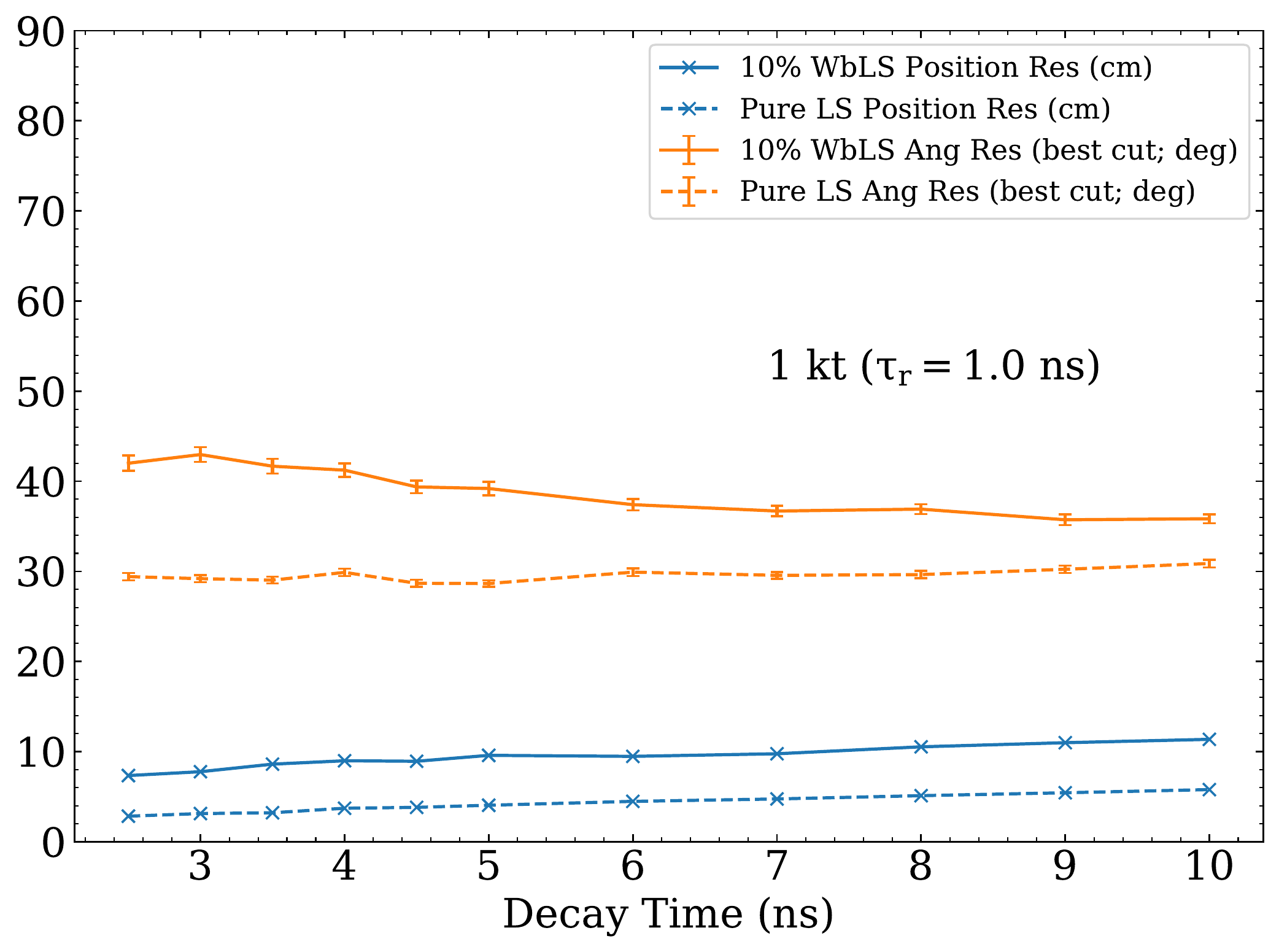} \\
\includegraphics[width=\columnwidth]{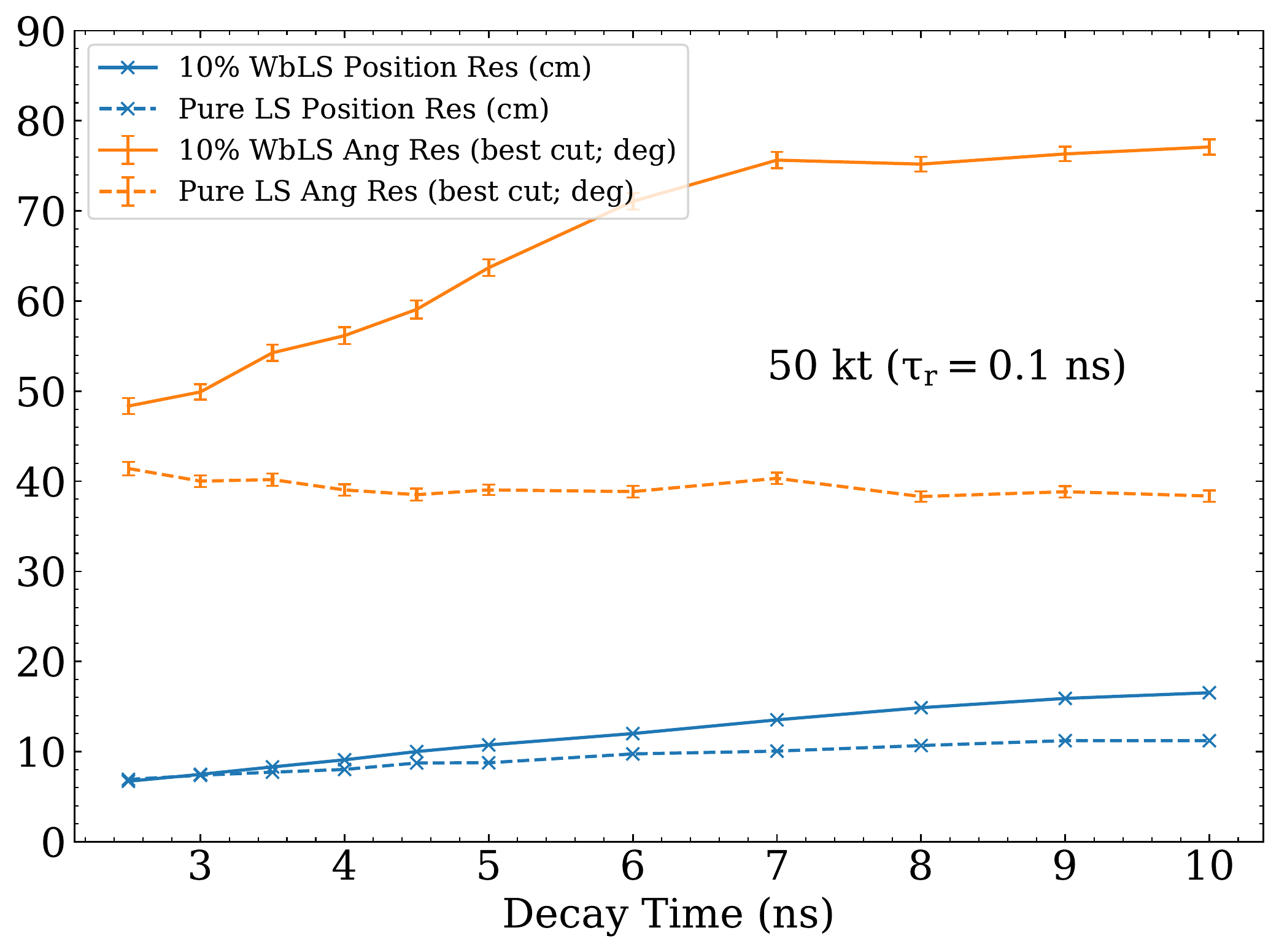}
\includegraphics[width=\columnwidth]{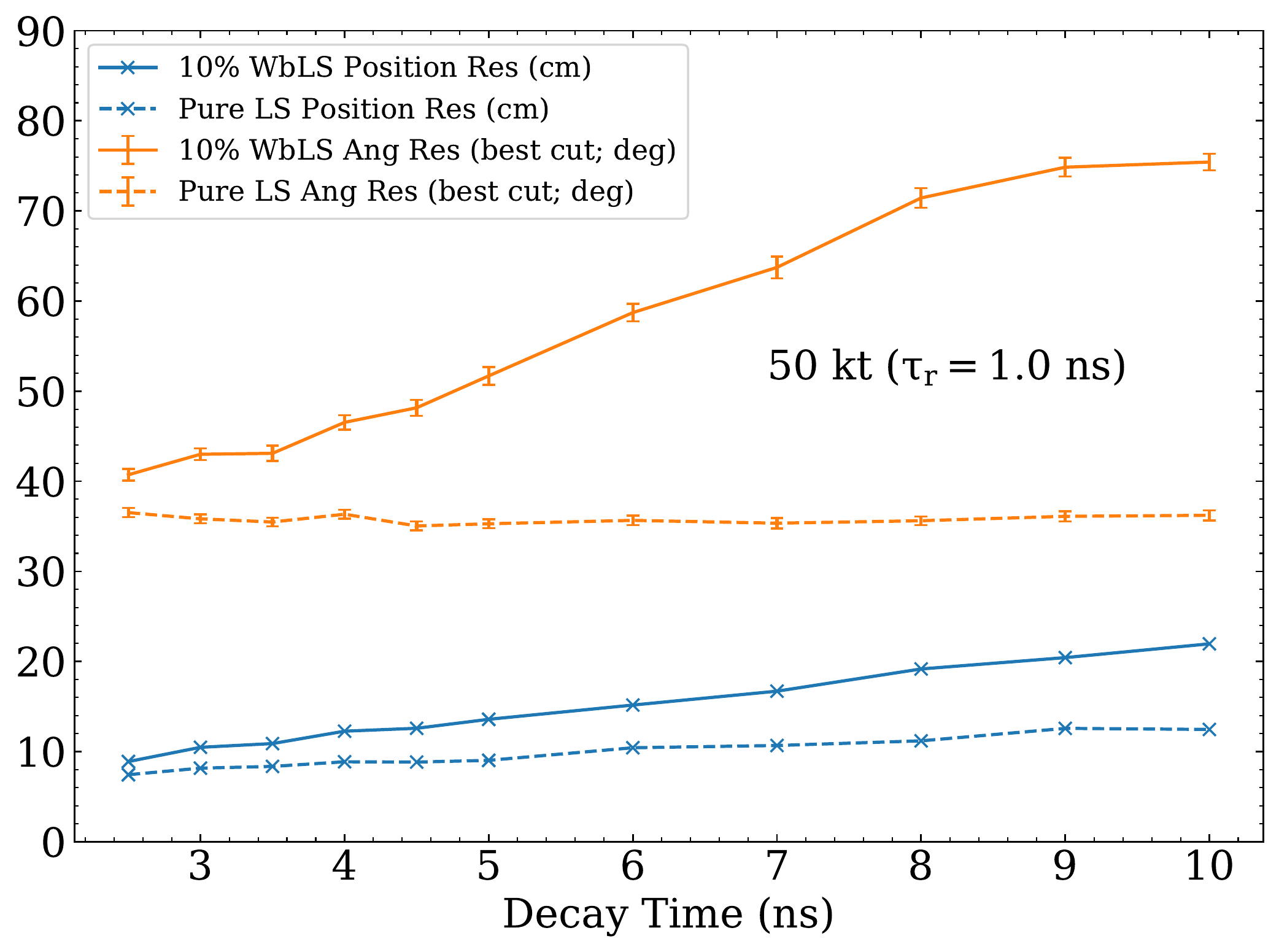} \\
\caption{\label{fig:decay-scan}Reconstruction resolutions for a scan of the scintillation decay time with a rise time of (left) 100~ps and (right) 1~ns in the (top) 1-kt detector geometry and (bottom) 50-kt detector geometry. Results are shown for the LAPPD photon detector model for the 10\% WbLS and pure LS materials. Angular resolution is shown for the best $t_{prompt}$ cut (see \Cref{app:cuts}). See legend for units.}
\end{figure*}

As before, this is done for 2.6-MeV electrons with both the 1-kt and 50-kt detector geometries.
Only the LAPPD photon detector model is explored here, to simplify the presentation of results.
Resolution metrics are presented for position and direction with the 10\% WbLS and pure LS materials in \Cref{fig:decay-scan}.  Energy resolution is unaffected by changes to the time profile.

Slower decay constants in 10\% WbLS appear to improve angular resolution quite significantly in the 1-kt geometry, more so for the faster rise time, but degrade the resolution in the 50-kt geometry.
The primary difference between these two geometries (for the same material and time profile) is the impact of dispersion (see \Cref{fig:hit-time-residuals}).
In the 50-kt geometry, there is a dispersion-dominated population of prompt Cherenkov photons independent of the time profile used.
Increasing the decay constant of the scintillator in this limit primarily broadens the time profile, which degrades the reconstruction metrics.
In the 1-kt geometry, which is not dominated by dispersion, the broadening of the time profile due to increasing decay constant does reduce the prompt scintillation light, resulting in improved angular reconstruction. 
This improvement is less significant with the larger rise time, as the larger rise time itself removes much of the prompt scintillation light.

Notably for pure LS the effects are small: slowing the scintillation light without modifying other parameters in the pure LS has little time impact on detector performance.
This indicates that the slower time profile of pure LS relative to WbLS is not the driving factor behind its good performance in these metrics, which is instead dominated by the impact of dispersion due to the high refractive index.

\subsection{Rise time}

\begin{figure*}[]
\centering
\includegraphics[width=\columnwidth]{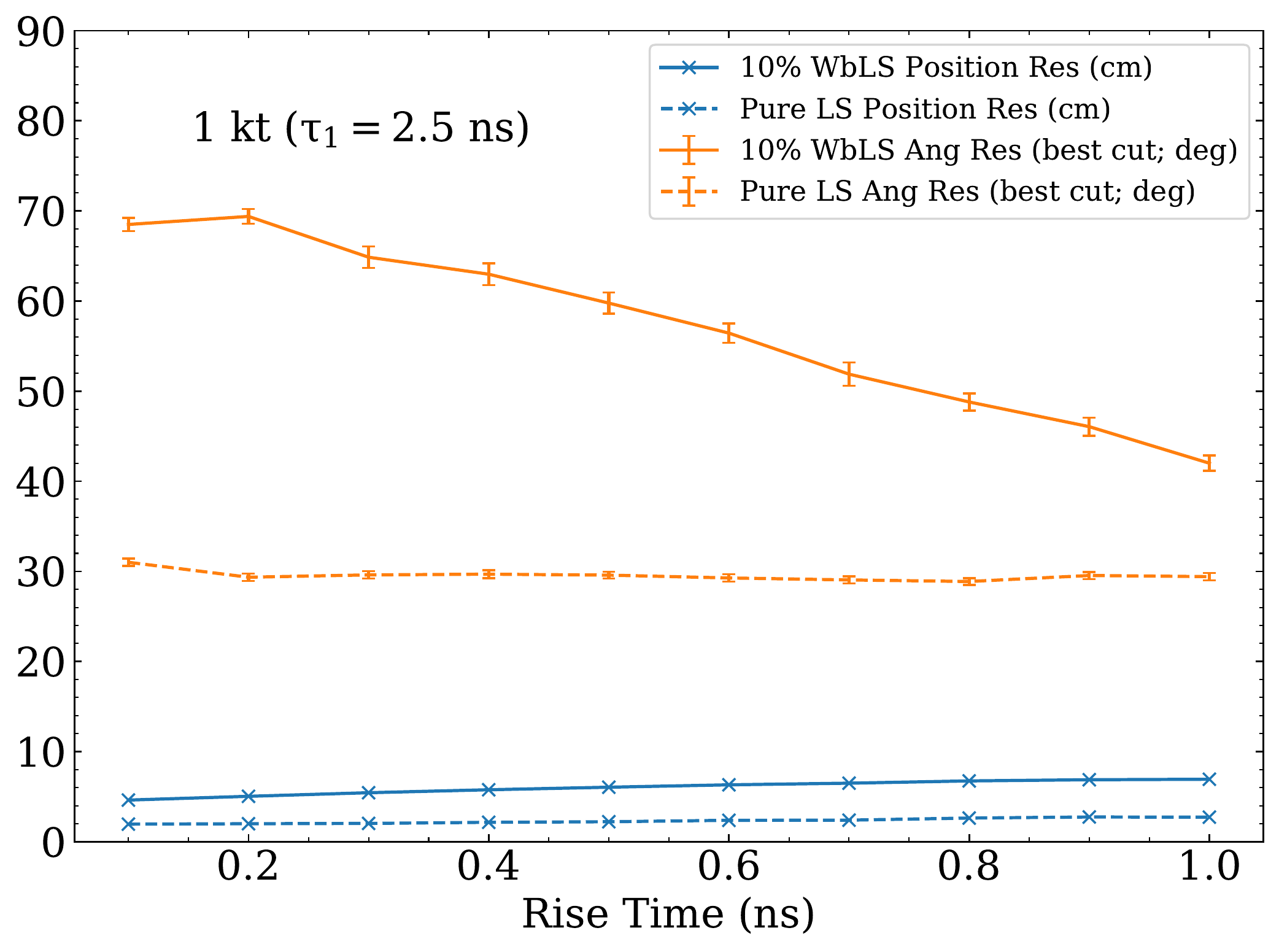} 
\includegraphics[width=\columnwidth]{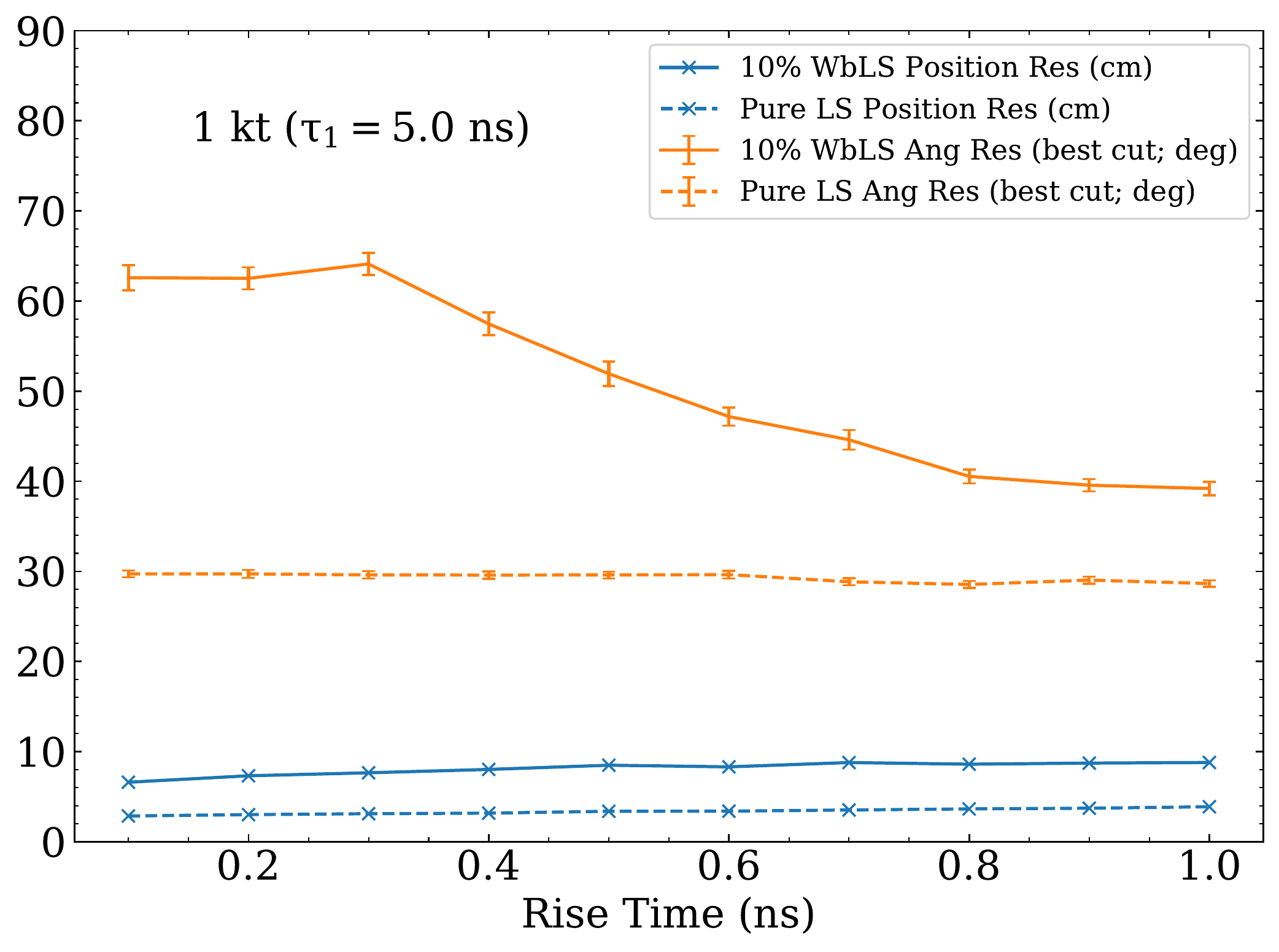} \\ 
\includegraphics[width=\columnwidth]{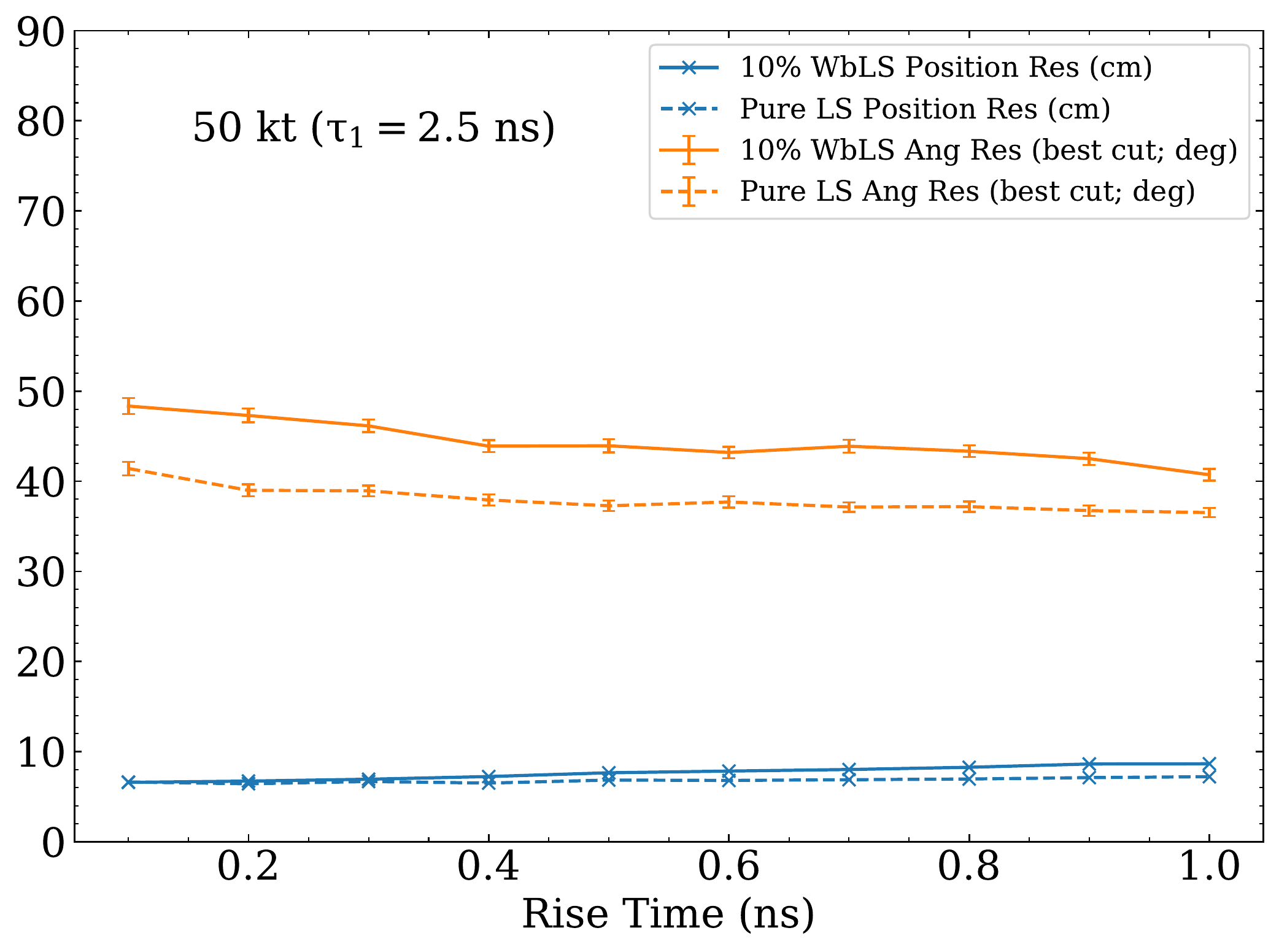} 
\includegraphics[width=\columnwidth]{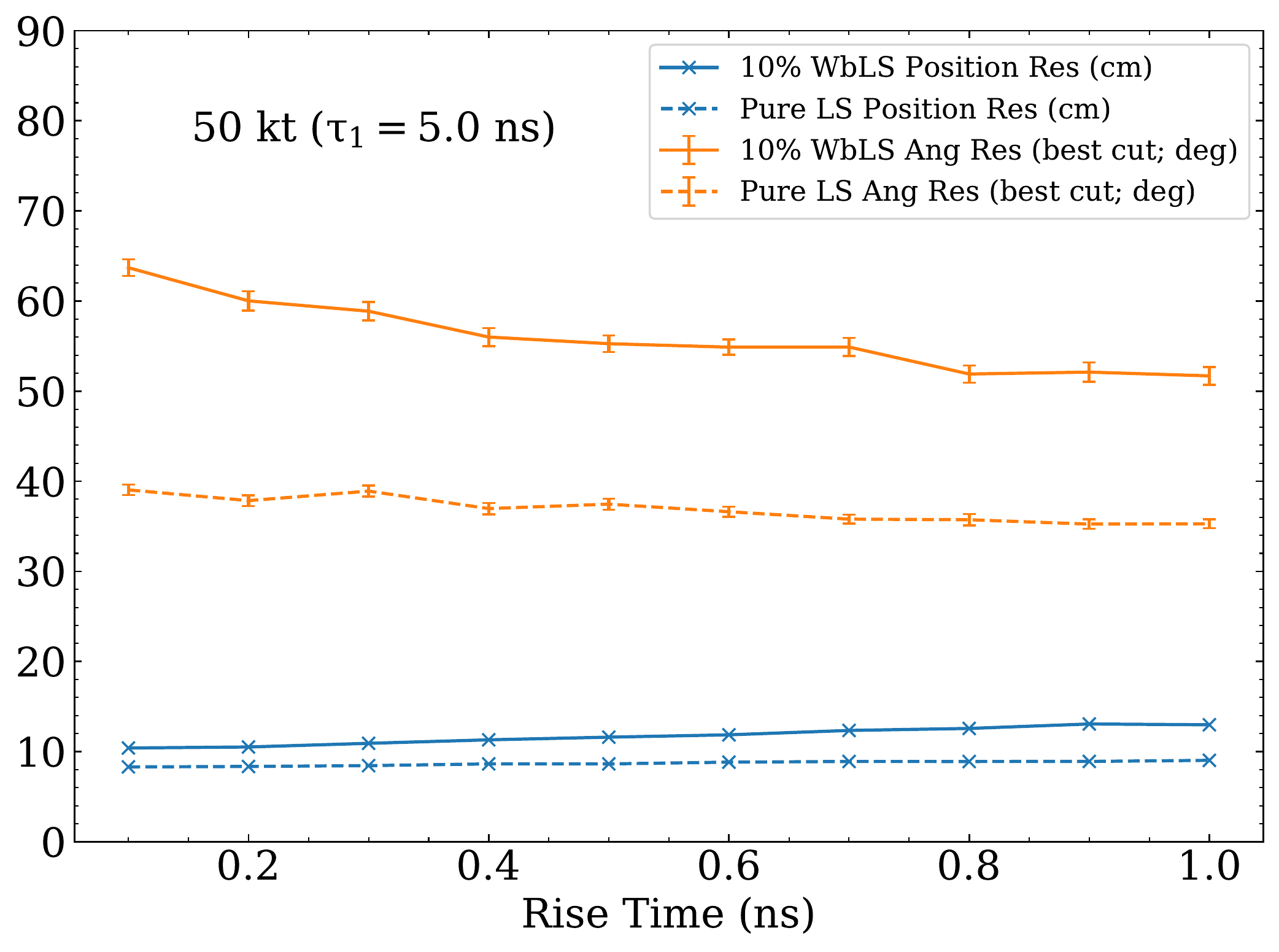} \\ 
\caption{\label{fig:rise-time-scan}Reconstruction resolutions when the scintillation rise time is scanned for a decay time of (left) 2.5 ns and (right)  5.0 ns in the (top) 1-kt detector geometry and (bottom) 50-kt detector geometry. This is done using the LAPPD photon detector for the 10\% WbLS and pure LS materials. Angular resolution is shown for the best $t_{prompt}$ cut (see \Cref{app:cuts}). See legend for units.}
\end{figure*}

Since increasing the decay time constant to spread out the scintillation light had adverse effects in the 10\% WbLS at the 50-kt detector, a scan of the rise time is performed to understand the impact on the reconstruction metrics.
The rise time is scanned for values from 100~ps to 1~ns, for both a 2.5~ns and 5~ns decay time, characteristic of WbLS and pure LS, respectively.  As before, this is done for 2.6-MeV electrons with both the 1-kt and 50-kt detector geometries.  Results are shown in \Cref{fig:rise-time-scan}.

In all cases, slowing the rise time improves the angular resolution, but slightly degrades the position and time resolution.
Slower rise times in 10\% WbLS degrade the position and time resolution more than in the pure LS material.
10\% WbLS demonstrates significant gains in angular resolution for slower rise time constants, and this is most pronounced in the 1-kt detector where the prompt Cherenkov is not yet well separated by dispersion.
Pure LS results in the best overall resolution, and is again minimally impacted by adjusting its time profile.
Simulated hit time residuals in \Cref{fig:hit-time-residuals} show that the unmodified pure LS material has a clear prompt Cherenkov population in the 50-kt detector (c.f. 10\% WbLS), which is not impacted significantly by adjusting the scintillation time profile.
This prompt Cherenkov population is the dominant factor in the good performance of pure LS compared to 10\% WbLS, and is primarily due to the greater impact of dispersion in pure LS.

\section{\label{sec:physics} Impact for physics reach}

We now briefly examine how the energy and angular resolutions evaluated in the previous sections affect the capability for rejection of the ${}^8$B solar neutrino background in NLDBD searches, and identification of signal events for CNO solar neutrino detection.  In both cases, identification (as either signal or background) of the directional solar neutrino events is the capability under study. 

Detailed studies have been performed in~\cite{theia_wp} of the sensitivity of a 50-kton (Wb)LS detector to both CNO neutrinos and to NLDBD.  However, in that paper a number of simplifying assumptions were made, including an assumed vertex and angular resolution, and simplified approach to energy reconstruction.  In addition, that work was based on previously understood, now outdated, properties for WbLS.  This work represents the first study using a data-driven optical model for WbLS, a more realistic detector simulation at the single photon level, and full event reconstruction.  This work therefore serves to validate the simpler assumptions made in~\cite{theia_wp} and to support the results from that work.

In order to do so, we again make use of the RAT-PAC framework~\cite{ratpac}, including the neutrino-electron elastic scattering generator and the radioactive decay generator used by SNO~\cite{sno_reconstruction} and SNO+~\cite{snoplus_nd} as well as an implementation of Decay0~\cite{Decay0lg}.  
In simulation, the neutrino-electron elastic scattering differential cross section~\cite{Bahcall:1995mm} is weighted by the neutrino energy spectrum~\cite{Winter:2004kf,Bahcall:1987jc} for the different fluxes from the Sun and then sampled in outgoing electron energy and scattering angle, for both $\nu_e$ and $\nu_\mu$.  Solar neutrino fluxes are taken from~\cite{Bahcall:2004pz}.
The decay energy spectra are also found for various backgrounds associated with the CNO energy region of interest. The solar neutrino interactions and decays are then simulated accordingly to extract the expected energy deposition in the target materials under consideration. After the simulation, solar neutrino event samples are weighted following the survival probability calculated in \cite{Bonventre:2013loa}.

The extracted angular resolution parameters from Secs.~\ref{sec:impact} and~\ref{sec:timing-impact} are used to smear the scattering angle for solar neutrino events using a functional form taken from~\cite{theia_richie}, while radioactive and cosmogenic background events, as well as double beta decay events, are assumed to be isotropic.  

\subsection{NLDBD sensitivity}

For the NLDBD study, we consider LAB+PPO loaded with 5\% natural Te (34.1\% $^{130}$Te), and assume the expected $3\% / \sqrt{E}$ energy resolution from~\cite{theia_wp}, since the isotope-loaded scintillator will behave differently from those studied here.  We intentionally make the same assumptions as in that previous work in order to do a direct comparison with the implementation of the more complete optical model and reconstruction presented here.
We make the same assumptions about location and background rates as in the previous study~\cite{theia_wp}, which should be referred to for further detail. Notably, ${}^8$B solar neutrino events are the dominant background.  The purpose of this study is to explore the impact of the angular resolutions determined in Sec.~\ref{sec:impact}.
No assumption on angular resolution was directly made in~\cite{theia_wp}, so we use the angular resolution found here for unloaded scintillator to extend the previous analysis, as being representative of reasonably achievable time profiles.  
Energy cuts are applied to restrict the study to the $0\nu\beta\beta$ region of interest for ${}^{130}$Te, as outlined in \cite{theia_wp}.  We further apply cuts as a function of reconstructed direction relative to the Sun,  
$\cos\theta_\odot$, in order to reduce the background from directional $^8$B solar neutrinos.
The fraction of $\nu_e$ and $\nu_\mu$ samples for ${}^8$B neutrinos surviving these analysis cuts are scaled according to expected event rates on LAB+PPO in order to maintain the correct ratio of $\nu_e$ and $\nu_\mu$ interactions and properly calculate the overall efficiency for rejecting solar neutrino background events and accepting isotropic events such as radioactive decays or $0\nu\beta\beta$. 

The efficiencies for the cut values are then propagated through the box  analysis procedure of \cite{theia_wp} to select an optimal cut that yields the best sensitivity. To quote an example, we find an expected sensitivity of $T^{0\nu\beta\beta}_{1/2} > 1.4 \times 10^{28}$ years at 90\% CL in the 50-kt, LAPPD-instrumented pure LAB+PPO detector with decay time of $2.5$ ns and rise time of $1.0$ ns, after 10 years of data taking.   This equates to a mass limit of $m_{\beta\beta} < 4.5-11.1$~meV, using nuclear matrix elements from~\cite{Rodriguez:2010mn,Barea:2013bz}. KamLAND-Zen has placed a limit on the effective Majorana neutrino mass of $61-165$~meV~\cite{KamLAND-Zen}, and the SNO+ experiment projects a sensitivity of $55-133$~meV~\cite{snoplus}.  Fig. 19 of \cite{theia_wp} shows this result in the context of other proposed future experiments.  Such a detector achieves an angular resolution of roughly 37${}^\circ$. This result is achieved by cutting on a solar angle corresponding to $\cos\theta_\odot = 0.7$, which rejects over 65\% of the $^{8}$B background while keeping 85\% of the signal. This increases confidence in assumptions of rejection capability used in \cite{theia_wp}. Notably, improving the angular resolution to 30${}^\circ$ and performing the same analysis does not yield changes to sensitivity to the leading decimal. Note that this result confirms that of more sophisticated reconstruction techniques, such as that presented in~\cite{Jiang:2019cnb}, in which similar rejection was demonstrated for a 3-m radius detector.  In this case we demonstrate that such rejection can be preserved even in the much larger detector under consideration here, which is critical for next-generation NLDBD sensitivity.

\begin{figure}[]
\centering
\includegraphics[width=\columnwidth]{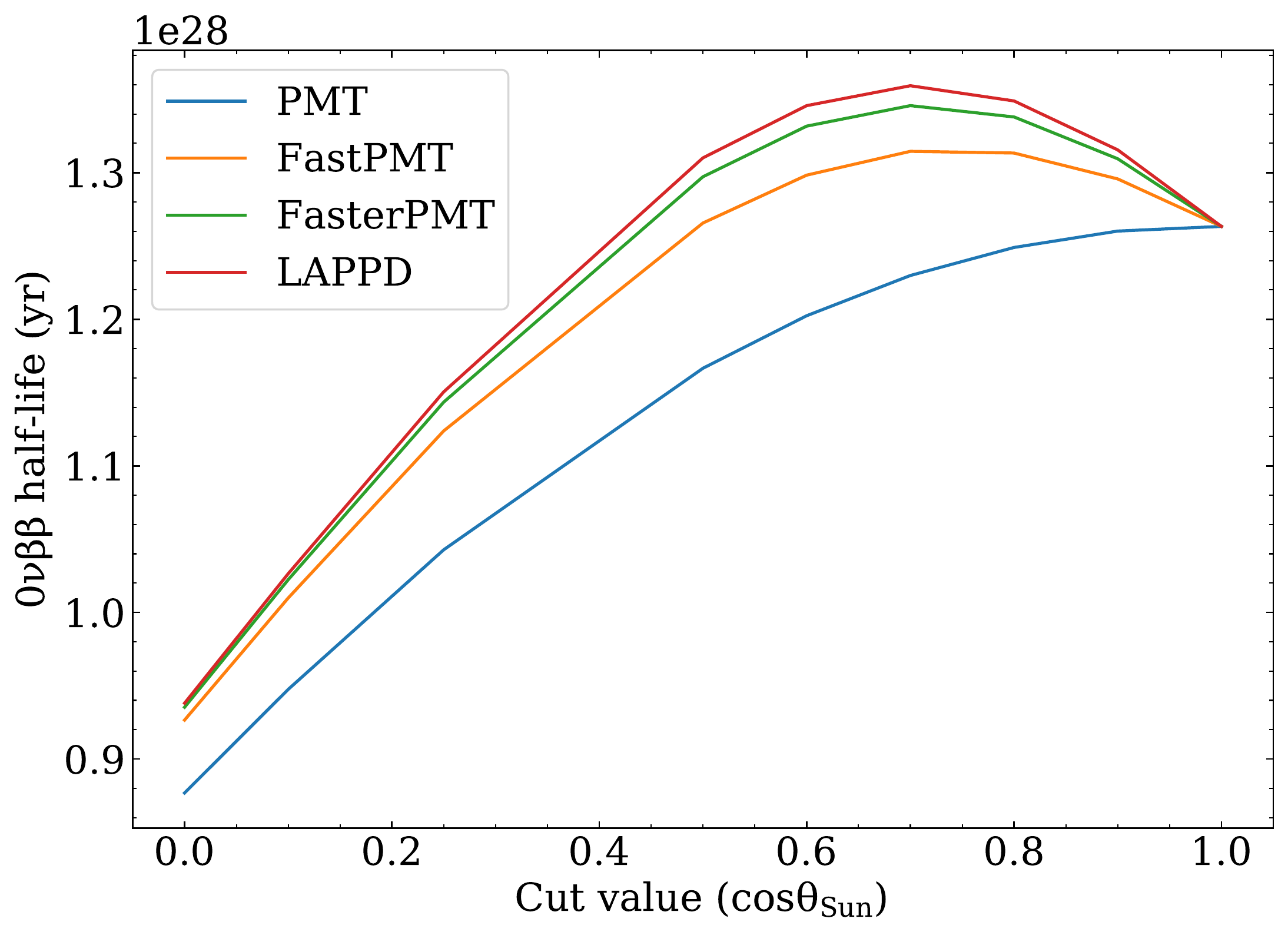} \\ 
\caption{\label{fig:DBD} Half-life sensitivity for $0\nu\beta\beta$ achieved for a 50-kt pure LS detector with an 8m radius balloon of Te-loaded pure LS at 5\% loading, as a function of solar angle cut and photodetector model. Angular resolution is based on that found in Sec.~\ref{sec:impact}, assuming the as-measured properties of LAB+PPO without considering possible delays to the scintillation profile, and we use 3\%/$\sqrt{E}$ energy resolution, as assumed in \cite{theia_wp}.}
\end{figure}

Several other configurations for the 50-kt detector give results with similar sensitivity. Fig.~\ref{fig:DBD} shows the impact of the various photon detector models, with only small losses in sensitivity for the 500-ns (FasterPMT) and 1-ns (FastPMT) models, of less than 1\% and approximately 3\% in lifetime, respectively.  Only standard PMTs show a significant degradation of sensitivity, and this detector is also seen to perform best with no cut on solar angle, due to the degraded direction resolution achieved for this configuration.  For the LAPPD-instrumented detector, we see that the impact of scanning the decay time for values from 2.5 to 10~ns for LAB+PPO changes the sensitivity by less than $0.02 \times 10^{28}$ years, and the sensitivity improves for slower rise times, but the impact of the change from a rise time of 100~ps to 1~ns is less than $0.04 \times 10^{28}$ years.
As such, variation of the decay and rise time of the scintillation time profile at the scale examined, without other changes to LS optical properties, are not thought to have a large impact on sensitivity to NLDBD.  It should be noted that this conclusion is specific to our particular choice of direction reconstruction methodology, and conclusions may differ for other approaches. 

\subsection{Precision CNO measurement}

We also evaluate scenarios for CNO solar neutrino detection in a manner akin to the large-scale WbLS detector studies presented in \cite{theia_richie} and \cite{theia_wp}.  We make the same assumptions about location and background rates as in those studies and, as in the NLDBD case, further details can be found therein.
Instead of the hit-based lookup reconstruction scheme applied in those studies, we employ a Gaussian smearing based on the expected number of hits, as determined in Sec.~\ref{sec:impact}.  Since quenching effects are fully simulated, we take only the part of the width that is due to photon counting, so as not to double count that effect.  The resolution is scaled with energy according to photon statistics.
The rest of the fitting procedure remains the same as that described in the mentioned analyses, though we consider the use of a constraint on the $pep$ flux at 1.4\% from the global analysis of \cite{Bergstrom:2016cbh}, which leverages the information afforded by the full $pp$-chain and solar luminosity on experimental data. Application of this constraint follows the methodology of the recent Borexino discovery~\cite{Agostini:2020lci,Agostini:2020mfq}.

Since the angular resolution evaluated at 2.6 MeV is expected to be much finer than at energies more relevant to the CNO search, for this study, we instead use resolution values determined using simulated electrons at 1.0 MeV. For consistency, the energy resolution is also recalculated at 1.0 MeV. At this energy, we find that in the 50 kt, LAPPD-instrumented detector, the angular resolution achieved by the fitter is $70^\circ$ for 1\% WbLS and $65^\circ$ for LAB+PPO, as opposed to $40^\circ$ and $36^\circ$ respectively at 2.6 MeV. The energy resolution is assumed to vary $\propto 1/\sqrt{E}$ and the angular resolution is assumed to be flat.  This does not fully incorporate expected improvements in resolution at higher energies, and degradation at lower energies.  A more sophisticated study implementing the full energy dependence is underway.  This result is intended to guide the reader as to the capabilities of this style of detector.
 Energy cuts are applied to the CNO solar neutrino fit region, following the approach in~\cite{theia_wp}.  We consider a threshold of 0.6~MeV in all cases.

It is of interest to see the direction reconstruction performance at these energies, with the acknowledged caveat that improvements are likely possible with more sophisticated analysis techniques.  Appendix~\ref{app:res} lists the direction resolution achieved for both the 1- and 50-kt detectors, for each target material, with each photon detector model, at both 1~MeV and 2.6~MeV.

\begin{figure}[]
\centering
\includegraphics[width=\columnwidth]{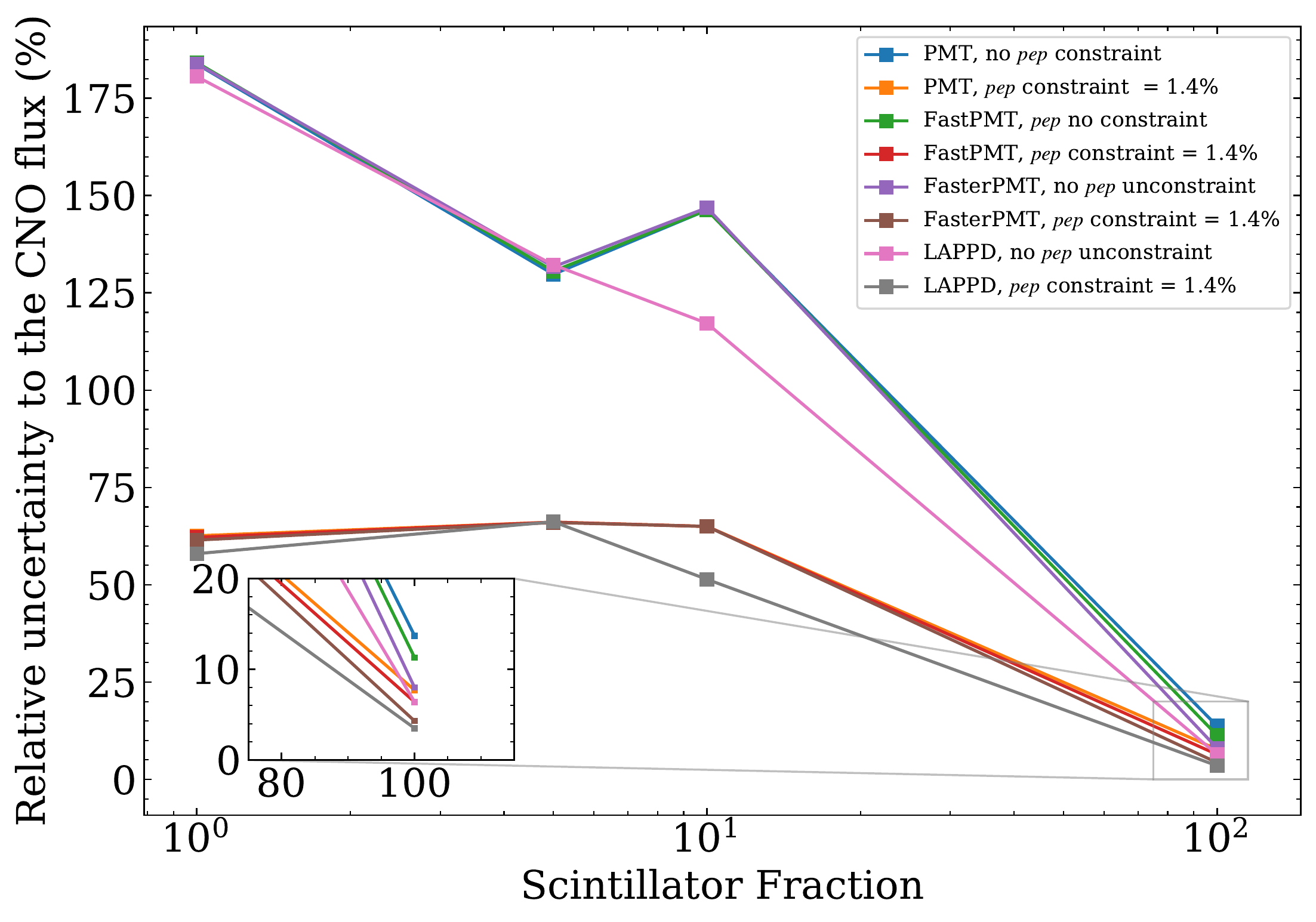} \\ 
\includegraphics[width=\columnwidth]{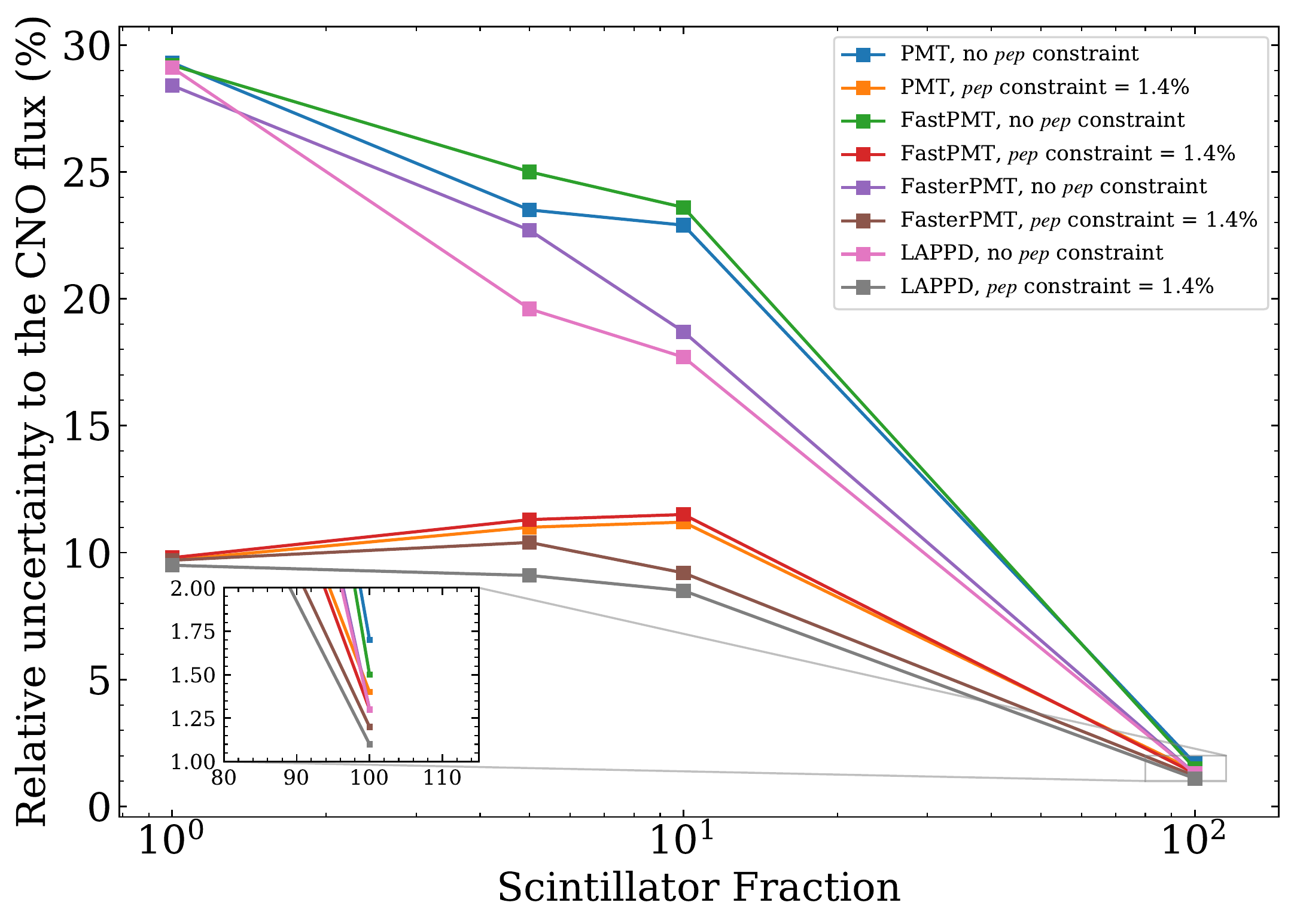} \\ 
\caption{\label{fig:CNO}(Top) Precision achieved for a measurement of the CNO flux in a 1-kt detector, as a function of the percentage of LS in the target material, where a value of 10$^2$ refers to pure LS, and of the photodetector model.  Detector performance is based on that found in Sec.~\ref{sec:impact}, assuming the as-measured properties of WbLS and LS, without considering possible delays to the scintillation profile.  The angular resolution and energy resolution have been recalculated at 1~MeV, according to the methodology outlined in earlier sections.
The inset shows a  zoom in on the pure LS sensitivity for the 1-kt detector, to illustrate the importance of photon detector model for this configuration.
(Bottom) CNO precision in the 50-kt detector, as a function of \%LS and photodetector model.}
\end{figure}

Fig.~\ref{fig:CNO} shows the results for the precision with which the CNO flux could be determined, in both the 1- and 50-kt detectors, for each combination of target material and photodetector model.  The 1-kt results are seen to have  little dependence on TTS for a WbLS deployment.  Due to the small target mass (500-ton fiducial volume, after a 50\% cut to reject external events) the sensitivity is significantly reduced in this smaller detector, and the dependence on target material is notably stronger, due to the reduced impact of dispersion for the shorter path lengths.  However, a pure LS detector can still achieve an excellent measurement of CNO neutrinos, with  dependence on photodetector model, due to the impact of direction resolution on background rejection efficiency.  Better than 5\% can be achieved in an LAPPD-instrumented detector.
In the 50-kt detector a stronger dependence on TTS is observed across the spectrum of target materials, although the achievable sensitivities are reasonably comparable across different photodetector models, with the largest variations observed for 5\% and 10\% WbLS, where tradeoffs between angular resolution and light yield become important.

We find that in 5 years of data taking, the CNO flux could be determined to a relative uncertainty of 18\% (8\%) in the 50-kt, LAPPD-instrumented 10\% WbLS detector when the $pep$ flux is unconstrained (constrained to 1.4\%), and to 1\% in the same detector filled with LAB+PPO, with the $pep$ flux either constrained or unconstrained. By contrast, Borexino's discovery includes a 1$\sigma$ uncertainty of 42\% above and 24\% below their measured flux, including statistical and systematic uncertainties~\cite{Agostini:2020mfq}.  We note that the result for the $pep$-constrained case is not very sensitive to the fraction of scintillator in WbLS (1--10\% perform similarly) whereas in the $pep$-unconstrained case the performance degrades with reduced scintillator fraction.  This is understood because the angular resolution is found to be similar for different WbLS materials at 1~MeV (approximately 70$^{\circ}$), so the light yield becomes the critical component in determining performance.
A more comprehensive study of these effects will be forthcoming in a future publication.

\section{\label{sec:conclusions} Conclusions}

In this paper we have considered the low-energy performance of both 1- and 50-kt detectors, with a range of target materials.  We focus on new measurements of WbLS, and their impacts on detector performance, but consider both pure water and pure scintillator detectors for comparison.  We also consider the impact of slowing the scintillation light in both the pure LS and the WbLS.  We consider four models for photon detectors, with time resolution of 1.6~ns, 1~ns, 500~ps, and 70~ps.  We study detector performance in terms of energy, vertex, and angular resolution, and go on to the interpret the results in terms of sensitivity to the CNO solar neutrino flux, and a search for NLDBD.

While LS outperforms WbLS for these particular physics goals, many factors motivate the choice of target material for a particular detector.  A large-scale WbLS detector would preserve a long-baseline program, offering similar sensitivity to neutrino mass hierarchy and CP violation as an additional DUNE module~\cite{theia_wp}, along with a broad program of low energy physics.  Other factors to consider include practical considerations such as cost, risk, deployment procedures, and purification and recirculation requirements.  In this paper we consider some of the potential physics and performance trade offs between such a large-scale WbLS deployment, a standard water Cherenkov detector, and a pure LS fill, and explore how these trade offs change across parameter space.

Different optical properties dominate many of the effects under consideration. Due to the higher refractive index, more Cherenkov photons are generated in pure scintillator than in water or WbLS, which competes with increased absorption and scattering in this material.  Effects of absorption and reemission can be seen in the large detector, where more reemitted photons are detected than direct scintillation photons.  

We evaluate energy resolution using the width of the detected hit distribution.  As expected, this increases with fraction of scintillator in the target, with minimal impact from the photon detector model.
We employ a likelihood-based evaluation of vertex and direction reconstruction.
The scintillation component of WbLS improves the vertex resolution but degrades the angular resolution relative to pure water.
The faster time profile of WbLS compared to pure LS makes the identification of the Cherenkov population more challenging, thus hindering direction reconstruction.  

Dispersion effects play a  significant role in the ability to separate Cherenkov photons, particularly in the larger detector.  We see that the impact of faster timing photon detectors on low-energy reconstruction performance is important in the larger detector size in order to fully leverage this effect for reconstruction.
The higher refractive index of pure LS increases the effects of dispersion for this material.  The optimal low-energy angular resolution in a scintillating detector is achieved for pure LS, under the assumption of 70-ps time resolution. For time resolutions of 1~ns or worse, water and WbLS perform better.   The difference in performance between WbLS and pure LS is much less significant in the larger detector, where 5\% and 10\% WbLS perform similarly to pure LS.  
It is worth noting that studies of direction reconstruction at high energies may yield different conclusions, given much higher photon statistics.

The fast time profile of WbLS motivated consideration of delaying the time profile, to understand the impact on detector performance.  Slow scintillators are under active development, in part for their potential to offer improved angular resolution for low-energy events.  This possibility was studied for both 10\% WbLS and for pure LS.   We observe minimal impact on either position or direction reconstruction for pure LS, but the angular resolution of WbLS can be significantly improved by slowing the scintillation light, to that equivalent to pure LS or even slower, with relatively small impact on vertex resolution.

We consider the impact of the observed detector performance for both CNO solar neutrino detection, and potential for deployment of a containment vessel of Te-loaded pure LS in a larger WbLS detector, for a search for Majorana neutrinos via NLDBD.  
We find that the 50-kt detector has sensitivity to the CNO neutrino flux of better than 20\% under conservative assumptions with no constraint on the pep flux, better than 10\% in a lightly loaded WbLS detector when considering a constraint on the pep flux, as was done for the recent Borexino discovery~\cite{Agostini:2020mfq}, and 1\% for a pure LS detector. A 1-kt total mass detector has reduced sensitivity due to the reduced statistics, but a pure LS deployment can still achieve a sub 5-percent measurement.
For NLDBD we find a half life sensitivity of $T^{0\nu\beta\beta}_{1/2} > 1.4 \times 10^{28}$ years at 90\% CL for 10 years of data taking, which equates to a mass limit of $m_{\beta\beta} < 4.5-11.1$~meV.  These results both have a weak dependence on photon detector model, with only small degradation in sensitivity for TTS values up to 1 ns.

\begin{acknowledgements}

The authors would especially like to thank Tanner Kaptanoglu, Michael Wurm, Josh Klein, Bob Svoboda, Matthew Malek and Adam Bernstein for useful discussions.  The authors would like to thank the SNO+ collaboration for providing a number of simulation modules, as well as data on the optical properties of the pure LS, including the light yield, absorption and reemission spectra, and refractive index.

This material is based upon work supported by the U.S. Department of Energy, Office of Science, Office of High Energy Physics, under Award Number DE-SC0018974. Work conducted at Lawrence Berkeley National Laboratory was performed under the auspices of the U.S. Department of Energy under Contract DE-AC02-05CH11231. The work conducted at Brookhaven National Laboratory was supported by the U.S. Department of Energy under contract DE-AC02-98CH10886. The project was funded by the U.S. Department of Energy, National Nuclear Security Administration, Office of Defense Nuclear Nonproliferation Research and Development (DNN R\&D).

\end{acknowledgements}

\bibliographystyle{apsrev4-1}
\bibliography{WbLS_extrap}  

\appendix

\section{Best $t_{prompt}$ cut values}
\label{app:cuts}

The best $t_{prompt}$ cuts for the results in this paper are reported here.
The $t_{prompt}$ cut was scanned from -1~ns to 5~ns in 0.25~ns steps, and from 5~ns to 10~ns in 1~ns steps.
The value that resulted in the smallest angular resolution was chosen as the best. 
Note that prompt cuts were not benificial to many conditions (seen as a $t_{prompt}$ of 10~ns here), but were especially useful in the case of very fast timing (LAPPD), materials with a great deal of dispersion (pure LS), or materials with slow rise and decay constants.
The $t_{prompt}$ values are shown for the scintillator fraction study in \Cref{tbl:scintfrac_cuts}, for the decay time study in \Cref{tbl:decaytime_cuts}, and for the rise time study in \Cref{tbl:risetime_cuts}.

\begin{table*}
\begin{tabular}{l|l||r|r|r|r|r}
Size & Photodetector & Water cut (ns) & 1\% WbLS cut (ns) & 5\% WbLS cut (ns) & 10\% WbLS cut (ns) & Pure LS cut (ns) \\ \hline
50 kt & PMT & 6.00 & 8.00 & 10.00 & 10.00 & 0.00 \\
50 kt & FastPMT & 10.00 & 10.00 & 10.00 & 10.00 & 0.00 \\
50 kt & FasterPMT & 5.00 & 9.00 & 10.00 & 0.00 & 0.50 \\
50 kt & LAPPD & 2.25 & 9.00 & 0.25 & 0.00 & 0.00 \\
1 kt & PMT & 2.25 & 5.00 & 10.00 & 10.00 & 0.00 \\
1 kt & FastPMT & 6.00 & 2.00 & 9.00 & 10.00 & 0.00 \\
1 kt & FasterPMT & 0.50 & 1.00 & 9.00 & 10.00 & 0.00 \\
1 kt & LAPPD & 3.00 & 0.75 & 9.00 & 0.25 & 0.50 \\
\end{tabular}
\caption{
\label{tbl:scintfrac_cuts}
The best $t_{prompt}$ cut for each condition in the results of the scintillator fraction study, presented in \Cref{sec:recon_results}. 
}
\end{table*}

\begin{table*}
\begin{tabular}{l|l|l||r|r}
Size & $\tau_r$ (ns) & $\tau_1$ (ns) & 10\% WbLS cut (ns) & Pure LS cut (ns) \\ \hline
50 kt & 0.1 & 10.0 & 3.00 & 0.00 \\
50 kt & 0.1 & 9.0 & 10.00 & 0.00 \\
50 kt & 0.1 & 8.0 & 10.00 & 0.00 \\
50 kt & 0.1 & 7.0 & 9.00 & 0.00 \\
50 kt & 0.1 & 6.0 & 0.00 & 0.00 \\
50 kt & 0.1 & 5.0 & 0.00 & 0.00 \\
50 kt & 0.1 & 4.5 & 0.25 & 0.00 \\
50 kt & 0.1 & 4.0 & 0.00 & 0.00 \\
50 kt & 0.1 & 3.5 & 0.00 & 0.00 \\
50 kt & 0.1 & 3.0 & 0.00 & 0.00 \\
50 kt & 0.1 & 2.5 & 0.00 & 0.00 \\
50 kt & 1.0 & 10.0 & 1.50 & 0.50 \\
50 kt & 1.0 & 9.0 & 1.25 & 0.50 \\
50 kt & 1.0 & 8.0 & 1.00 & 0.25 \\
50 kt & 1.0 & 7.0 & 0.75 & 0.50 \\
50 kt & 1.0 & 6.0 & 1.00 & 0.50 \\
50 kt & 1.0 & 5.0 & 0.75 & 0.00 \\
50 kt & 1.0 & 4.5 & 0.50 & 0.25 \\
50 kt & 1.0 & 4.0 & 0.00 & 0.00 \\
50 kt & 1.0 & 3.5 & 0.50 & 0.00 \\
50 kt & 1.0 & 3.0 & 0.25 & 0.00 \\
50 kt & 1.0 & 2.5 & 0.25 & 0.25 \\
1 kt & 0.1 & 10.0 & 0.75 & 0.25 \\
1 kt & 0.1 & 9.0 & 0.75 & 0.00 \\
1 kt & 0.1 & 8.0 & 0.75 & 0.25 \\
1 kt & 0.1 & 7.0 & 0.75 & 0.00 \\
1 kt & 0.1 & 6.0 & 0.75 & 0.00 \\
1 kt & 0.1 & 5.0 & 0.75 & 0.00 \\
1 kt & 0.1 & 4.5 & 0.75 & 0.25 \\
1 kt & 0.1 & 4.0 & 10.00 & 0.25 \\
1 kt & 0.1 & 3.5 & 8.00 & 0.00 \\
1 kt & 0.1 & 3.0 & 7.00 & 0.00 \\
1 kt & 0.1 & 2.5 & 10.00 & 0.00 \\
1 kt & 1.0 & 10.0 & 0.50 & 0.25 \\
1 kt & 1.0 & 9.0 & 0.75 & 0.25 \\
1 kt & 1.0 & 8.0 & 0.75 & 0.50 \\
1 kt & 1.0 & 7.0 & 0.50 & 0.25 \\
1 kt & 1.0 & 6.0 & 0.50 & 0.50 \\
1 kt & 1.0 & 5.0 & 0.50 & 0.25 \\
1 kt & 1.0 & 4.5 & 0.50 & 0.50 \\
1 kt & 1.0 & 4.0 & 0.50 & 0.50 \\
1 kt & 1.0 & 3.5 & 0.75 & 0.25 \\
1 kt & 1.0 & 3.0 & 0.50 & 0.50 \\
1 kt & 1.0 & 2.5 & 0.50 & 0.25 \\
\end{tabular}
\caption{
\label{tbl:decaytime_cuts}
The best $t_{prompt}$ cut for each condition in the results of the decay time study, presented in \Cref{sec:timing-impact}. 
These cuts were found with the LAPPD photodetector model.
}
\end{table*}

\begin{table*}
\begin{tabular}{l|l|l||r|r}
Size & $\tau_r$ (ns) & $\tau_1$ (ns) & 10\% WbLS cut (ns) & Pure LS cut (ns) \\ \hline
50 kt & 0.1 & 2.5 & 0.00 & 0.00 \\
50 kt & 0.2 & 2.5 & 0.00 & 0.00 \\
50 kt & 0.3 & 2.5 & 0.00 & 0.00 \\
50 kt & 0.4 & 2.5 & 0.00 & 0.00 \\
50 kt & 0.5 & 2.5 & 0.00 & 0.00 \\
50 kt & 0.6 & 2.5 & 0.00 & 0.00 \\
50 kt & 0.7 & 2.5 & 0.00 & 0.00 \\
50 kt & 0.8 & 2.5 & 0.00 & 0.00 \\
50 kt & 0.9 & 2.5 & 0.25 & 0.25 \\
50 kt & 1.0 & 2.5 & 0.25 & 0.25 \\
50 kt & 0.1 & 5.0 & 0.00 & 0.00 \\
50 kt & 0.2 & 5.0 & 0.00 & 0.00 \\
50 kt & 0.3 & 5.0 & 0.25 & 0.00 \\
50 kt & 0.4 & 5.0 & 0.00 & 0.00 \\
50 kt & 0.5 & 5.0 & 0.00 & 0.25 \\
50 kt & 0.6 & 5.0 & 0.00 & 0.00 \\
50 kt & 0.7 & 5.0 & 0.50 & 0.00 \\
50 kt & 0.8 & 5.0 & 0.50 & 0.50 \\
50 kt & 0.9 & 5.0 & 0.75 & 0.00 \\
50 kt & 1.0 & 5.0 & 0.75 & 0.00 \\
1 kt & 0.1 & 2.5 & 10.00 & 0.00 \\
1 kt & 0.2 & 2.5 & 7.00 & 0.00 \\
1 kt & 0.3 & 2.5 & 0.25 & 0.25 \\
1 kt & 0.4 & 2.5 & 0.25 & 0.25 \\
1 kt & 0.5 & 2.5 & 0.25 & 0.00 \\
1 kt & 0.6 & 2.5 & 0.50 & 0.25 \\
1 kt & 0.7 & 2.5 & 0.50 & 0.25 \\
1 kt & 0.8 & 2.5 & 0.50 & 0.25 \\
1 kt & 0.9 & 2.5 & 0.50 & 0.25 \\
1 kt & 1.0 & 2.5 & 0.50 & 0.25 \\
1 kt & 0.1 & 5.0 & 0.75 & 0.00 \\
1 kt & 0.2 & 5.0 & 0.75 & 0.25 \\
1 kt & 0.3 & 5.0 & 0.75 & 0.25 \\
1 kt & 0.4 & 5.0 & 0.75 & 0.25 \\
1 kt & 0.5 & 5.0 & 0.75 & 0.25 \\
1 kt & 0.6 & 5.0 & 0.50 & 0.50 \\
1 kt & 0.7 & 5.0 & 0.75 & 0.50 \\
1 kt & 0.8 & 5.0 & 0.50 & 0.50 \\
1 kt & 0.9 & 5.0 & 0.50 & 0.50 \\
1 kt & 1.0 & 5.0 & 0.50 & 0.25 \\
\end{tabular}
\caption{
\label{tbl:risetime_cuts}
The best $t_{prompt}$ cut for each condition in the results of the rise time study, presented in \Cref{sec:timing-impact}.
These cuts were found with the LAPPD photodetector model.
}
\end{table*}

\section{Angular resolution}
\label{app:res}
Table~\ref{tab:resn} reports the achieved angular resolution for both the 1- and 50-kton detectors, for each target material, as a function of photon detector model, at both 1~MeV and 2.6~MeV.

\begin{table*}
\centering
\begin{tabular}{c|cc||cccc}
\multicolumn{1}{l}{} & \multicolumn{1}{l}{} & \multicolumn{1}{l||}{} & \multicolumn{4}{c}{Photodetector}                                              \\
Detector Size (kT)    & Energy (MeV)         & Material              & PMT   & FastPMT                      & FasterPMT & LAPPD                        \\ \hline
1                     & 1.0                  & Water                 & 38.5  & 38.2                         & 37.3      & 37.7                         \\
1                     & 1.0                  & 1\% WbLS              & 68.4  & 67.8                         & 67.3      & 64.6                         \\
1                     & 1.0                  & 5\% WbLS              & 85.5  & 85.6                         & 85.9      & 86.0                         \\
1                     & 1.0                  & 10\% WbLS             & 93.1  & 93.1                         & 92.7      & 74.8                         \\
1                     & 1.0                  & Pure LS               & 102.0 & 85.0                         & 58.8      & 44.8                         \\ \cline{2-7} 
1                     & 2.6                  & Water                 & 32.5  & 32.5                         & 32.6      & 32.4                         \\
1                     & 2.6                  & 1\% WbLS              & 38.4  & 37.3                         & 35.6      & 33.7                         \\
1                     & 2.6                  & 5\% WbLS              & 55.1  & 54.9                         & 54.5      & 54.2                         \\
1                     & 2.6                  & 10\% WbLS             & 68.2  & 68.0                         & 68.4      & 63.0                         \\
1                     & 2.6                  & Pure LS               & 89.5  & 62.7                         & 32.6      & 29.4                         \\ \hline
50                    & 1.0                  & Water                 & 44.9  & 43.0                         & 44.7      & 43.8                         \\
50                    & 1.0                  & 1\% WbLS              & 70.2  & 69.9                         & 70.1      & 69.9                         \\
50                    & 1.0                  & 5\% WbLS              & 86.7  & 86.3                         & 82.0      & 73.6                         \\
50                    & 1.0                  & 10\% WbLS             & 93.2  & 92.8                         & 78.8      & 71.8                         \\
50                    & 1.0                  & Pure LS               & 85.4  & 73.6                         & 67.7      & 64.8                         \\ \cline{2-7} 
50                    & 2.6                  & Water                 & 33.1  & 32.5                         & 33.0      & 33.0                         \\
50                    & 2.6                  & 1\% WbLS              & 40.4  & 38.4                         & 40.5      & 40.4 \\
50                    & 2.6                  & 5\% WbLS              & 56.5  & 55.1                         & 56.3      & 47.8                         \\
50                    & 2.6                  & 10\% WbLS             & 68.1  & 68.2                         & 53.0      & 44.7 \\
50                    & 2.6                  & Pure LS               & 58.5  & 89.5                         & 37.8      & 36.2                         \\ 
\end{tabular}
\caption{\label{tab:resn} The angular resolution in degrees selected with the best $t_{prompt}$ cut for each detector configuration explored.}
\end{table*}

\end{document}